\documentclass[acmsmall,screen]{acmart}

\setcopyright{none}

\newif\iflong
\newif\ifshort
\shortfalse
\longtrue

\usepackage[utf8]{inputenc}
\usepackage[T1]{fontenc}
\usepackage{xspace}
\usepackage{xparse}
  \NewDocumentCommand{\li}{v}{\textbf{\footnotesize\texttt{#1}}}
\usepackage[frozencache=true]{minted}
  \ifshort
  \setminted{fontsize=\footnotesize,linenos}
  \fi
  \iflong
  \setminted{fontsize=\footnotesize,linenos}
  \fi
\usepackage{tikz}
  \usetikzlibrary{calc}
  \usetikzlibrary{shadows,graphs}
\usepackage{amsmath}
\usepackage{mathpartir}
  
  \newcommand{\Rule}[1]{\hyperlink{#1}{\TirName {#1}}}

\newcommand{\sref}[1]{Section~\ref{sec:#1}}
\newcommand{\fref}[1]{Figure~\ref{fig:#1}}

\newcommand{\aeneas}{\textsc{Aeneas}\xspace}
\newcommand{\charon}{\textsc{Charon}\xspace}
\newcommand{\fstar}{F$^\ast$\xspace}

\DeclareRobustCommand{\aymeric}[1]\relax

\newcommand\kw[1]{\ensuremath{\mathsf{#1}}}
\newcommand\tbrw[2]{\ensuremath{\mathsf{\&}^#1\,#2}}
\newcommand\tmbrw[2]{\ensuremath{\mathsf{\&}^#1\mathsf{mut}\,#2}}
\newcommand\ebrw[1]{\ensuremath{\mathsf{\&}\,#1}}
\newcommand\embrw[1]{\ensuremath{\mathsf{\&mut}\,#1}}
\newcommand\eassign[2]{\ensuremath{#1 := #2}}
\newcommand\ederefs[1]{\ensuremath{*^s#1}}
\newcommand\ederefm[1]{\ensuremath{*^m#1}}
\newcommand\ederefb[1]{\ensuremath{*^b#1}}

\newcommand\emove[1]{\ensuremath{\kw{move}\,#1}}
\newcommand\ecopy[1]{\ensuremath{\kw{copy}\,#1}}
\newcommand\epanic{\kw{panic}}
\newcommand\ereturn{\kw{return}}
\newcommand\eseq[2]{#1;\,#2}
\newcommand\eite[3]{\kw{if}\,#1\,\kw{then}\,#2\,\kw{else}\,#3}

\newcommand\ematch[2]{\kw{match}\,#1\,\kw{ with }\;#2}
\newcommand\enone{\kw{None}}
\newcommand\esome[1]{\kw{Some}\;#1}
\newcommand\krv{\ensuremath{rv}}
\newcommand\kop{\ensuremath{op}}

\newcommand\kfalse{\mathsf{false}}
\newcommand\ktrue{\mathsf{true}}

\newcommand\emborrow[2]{\ensuremath{\mathsf{borrow}^m\,#1\;#2}}
\newcommand\esborrow[1]{\ensuremath{\mathsf{borrow}^s\,#1}}
\newcommand\eiborrow[1]{\ensuremath{\mathsf{borrow}^r\,#1}}

\newcommand\esloan[2]{\ensuremath{\mathsf{loan}^s\,\{#1\}\,#2}}
\newcommand\emloan[1]{\ensuremath{\mathsf{loan}^m\,#1}}
\newcommand\ebox[1]{\ensuremath{\mathsf{Box}\,#1}}

\newcommand\elproj[1]{\ensuremath{\kw{proj}_\mathsf{l}\,#1}}
\newcommand\ebproj[1]{\ensuremath{\kw{proj}_\mathsf{out}\,#1}}
\let\eoproj\ebproj
\newcommand\eiproj[1]{\ensuremath{\kw{proj}_\mathsf{in}\,#1}}
\newcommand\etproj[2]{\ensuremath{\kw{proj}_{#1}\,#2}}

\newcommand{\mcrot}[4]{\multicolumn{#1}{#2}{\rlap{\rotatebox{#3}{#4}~}}}

\ifshort
\newcommand\myparagraph[1]{\emph{#1}.\ }
\fi
\iflong
\let\myparagraph\paragraph
\fi

\newlength{\characterlength}
\newcommand\cspace{\hspace{\characterlength}}

\definecolor{envcolor}{RGB}{61, 122, 122}

\begin{document}

\title{\aeneas: Rust Verification by Functional Translation}

\author{Son Ho}
\affiliation{\institution{Inria}\country{France}}
\email{son.ho@inria.fr}

\author{Jonathan Protzenko}
\affiliation{\institution{Microsoft Research}\country{USA}}
\email{protz@microsoft.com}

\begin{abstract}
  We present \aeneas, a new verification toolchain for Rust programs based on a
  lightweight functional translation. We leverage Rust's rich region-based type
  system to eliminate memory reasoning  for a large class of Rust programs, as long
  as they do not rely on interior mutability or unsafe code. Doing so, we relieve the proof
  engineer of the burden of memory-based reasoning, allowing them to instead
  focus on \emph{functional} properties of their code.

  The first contribution of \aeneas is a new approach to borrows and
  controlled aliasing. We propose a pure, functional semantics for LLBC, a
  Low-Level Borrow Calculus that captures a large subset of Rust programs.
  Our semantics is value-based, meaning there is no notion of memory, addresses
  or pointer arithmetic. Our semantics is also ownership-centric, meaning that
  we enforce soundness of borrows via a semantic criterion based on \emph{loans}
  rather than through a syntactic type-based \emph{lifetime} discipline. We
  claim that our semantics captures the \emph{essence} of the borrow mechanism
  rather than its current implementation in the Rust compiler.

  The second contribution of \aeneas is a translation from LLBC to a pure
  lambda-calculus. This allows the user to reason about the original
  Rust program through the theorem prover of their choice, and fulfills our
  promise of enabling lightweight verification of Rust programs.
  To deal with the well-known technical difficulty of terminating a borrow, we
  rely on a novel approach, in which we approximate the borrow graph in the
  presence of function calls. This in turn allows us to perform the translation
  using a new technical device called \emph{backward functions}.

  We implement our toolchain in a mixture of Rust
  and OCaml; our chief case study is a low-level, resizing hash table, for which
  we prove functional correctness, the first such result in Rust.
  Our evaluation shows significant gains of verification
  productivity for the programmer. This paper therefore establishes a new point
  in the design space of Rust verification toolchains, one that aims to verify
  Rust programs simply, and at scale.

  Rust goes to great lengths to enforce static control of aliasing; the proof
  engineer should not waste any time on memory reasoning when so much already
  comes “for free”!
\end{abstract}




\maketitle

\section{Introduction}

In 2006, exasperated by yet another crash of his building's elevator's firmware, and
exhausted after walking up 21 flights of stairs, Graydon Hoare set out to design a
new programming language~\cite{rust-anecdote}. The language, soon to be
known as Rust, had two goals. First, to be system-oriented, meaning the
programmer would deal with references, pointers, and manually manage memory.
Second, to be safe, meaning the compiler's static discipline
would rule out memory errors such as use-after-free, or arbitrary memory access.
Even though the language evolved a great deal since its inception, these
two core premises remain today.

Sixteen years later, Rust enjoys a substantial amount of success, and has ranked as the most
loved programming language for six consecutive years on StackOverflow's
developer survey~\cite{stackoverflow}.
But as the systems community can
attest~\cite{klein2009sel4,lorch2020armada,ferraiuolo2017komodo,bhargavan2017everest},
memory safety is too weak of a property, no matter how remarkable of
an achievement Rust is.
Indeed, we oftentimes want to prove deep \emph{correctness properties}
of a system. Doing so may involve anything from baseline safety
properties, such as the absence
of assertion failures or runtime errors,
to complex invariants involving
concurrent systems.

As a consequence, several verification toolchains have emerged to facilitate
proving deep
properties about low-level programs. For pragmatic reasons, C is oftentimes a target of
choice~\cite{protzenko2017verified,cao2018vst}; so is using
a custom language, such as Dafny~\cite{hawblitzel2014ironclad}. Alas,
whether the tool is based on separation logic~\cite{reynolds2002separation} or
modifies-clauses~\cite{leino2010dafny}, verification
engineers soon find themselves drowning under a sea of memory-related
obligations that distract them from the properties of interest.
The net result is that verification engineers spend an undue amount of time
discharging mundane memory-related proof obligations, because the frontend language does not
enforce enough invariants to begin with.

One strategy is to write a checker for an existing language, restricting its usage
enough that verification becomes easier. This is the strategy used by e.g.,
Linear Dafny~\cite{lineardafny}
or RefinedC~\cite{sammler2021refinedc}.
Another strategy is to leverage invariants provided for free by languages with
restrictive type systems; that is, to leverage a language like Rust.

Rust is, in effect, trying to reconcile systems programming, and a long
tradition of static ownership disciplines~\cite{boyland2001capabilities,clarke1998ownership} which
traces back to linear types~\cite{wadler1990linear},
regions~\cite{tofte1997region} and the combination
thereof~\cite{fluet2006linear}. Today, Rust incorporates \emph{ownership} in all
aspects of its design; notably, programmers rely on \emph{borrowing} to take
references with ownership, and the type system relies on a notion of
\emph{lifetime} to enforce the soundness of borrows.  This static discipline aims
to give the programmer maximum flexibility for common idioms, while still
preventing arbitrary aliasing.

The adequacy of Rust as a verification target
 has not gone un-noticed. There
are now several research projects aiming to set up verification frameworks for
Rust, using a variety of backends, such as
SMT~\cite{matsushita2020rusthorn,verus},
Viper~\cite{prusti,prusti21}, or Why~\cite{creusot}.
With these works, we have gained a deeper understanding of why
verifying Rust programs remains difficult, even in the presence of Rust's strong ownership
discipline.

First, Rust's type system is not \emph{simply} linear, and features a rich
variety of mechanisms, such as: reborrows, two-phase borrows, functions returning borrows, along with many
other subtle rules that ensure most common idioms go through the type-checker
smoothly. Accounting for all of these is notoriously difficult. Second, the
borrow mechanism introduces non-locality, in that ending a borrow requires
propagating knowledge backwards, in order to update the previously-borrowed
variable with a new value. This central difficulty is
handled via a variety of technical devices in other works, such as prophecy
variables~\cite{matsushita2020rusthorn} or \li+after+ clauses~\cite{verus}.

In this paper, we propose a new approach to understanding and verifying Rust
programs. At the heart of our methodology is a lightweight functional
translation of Rust programs. We eschew the complexity of connecting to a
separation-logic based backend~\cite{jung2017rustbelt}, or relying on prophecy
variables to produce a logical
encoding~\cite{RustHornBelt, matsushita2020rusthorn}.
Instead, we synthesize a pure, functional, executable
equivalent of the original Rust program, thus producing a lambda-term that does
not rely on memory or special constructs. Our translation handles shared, mutable, two-phase and
re-borrows, and thus accounts for a very large fraction of typical Rust
programs.
We call the conceptual framework, as well as the companion tool, \aeneas.

We wish to emphasize that our functional translation is completely generic.
While we demonstrate a possible verification backend by printing our pure
programs in \fstar syntax, many other options are possible.  One could easily
add additional backends (Coq, Lean, Viper, Why) or devise a contract language
for source Rust programs that directly emits SMT proof obligations using our
translation. We certainly hope to write some of these in the near future.

To elaborate on the design choices made by \aeneas:
we intentionally focus on programs that
abide by Rust's static ownership discipline. That is, we do not tackle
\li+unsafe+ blocks -- we believe such programs are better suited to a
sophisticated framework such as RustBelt. We do not tackle interior mutability
either, in which the user can use unfettered aliasing, in exchange for a
run-time borrow checker.  We wish to focus instead on the Rust subset that
is functional in essence, meaning we leave treatment of interior mutability up
to future work. We believe this places us in a ``sweet spot'' for
verifying Rust programs.
The key observation of this work is that for the most part, references and borrows serve the
purpose of optimizing either performance (e.g., passing by reference instead
of by value), or memory representation (e.g., by controlling aliasing and taking
inner pointers within data structures).
That is, Rust's references do not serve
any semantic purpose; coupled with the fact that the type
system is
enforcing a linear discipline, such programs are functional in essence, and can be naturally translated to a pure
functional equivalent. Wadler observed that a linear type system allows
compiling pure programs using imperative updates~\cite{wadler1990linear}; we
leverage the reverse observation~\cite{chargueraud2008functional}, that is,
imperative programs with a strong enough ownership discipline admit
a functional equivalent.

To reiterate: the key point of this work is that we give a \emph{functional}
semantics, and thus a \emph{pure} translation, to the subset of Rust we
consider. Concretely, we define LLBC, the Low-Level Borrow Calculus, to
model that subset. Then, we give it an operational semantics that is
functional in nature. We do not rely on memory, addresses or pointer arithmetic;
rather, we map variables to values, and track aliasing in a very fine-grained
manner. We claim that our operational semantics captures the \emph{essence} of
borrowing; that is, it does not simply apply the rules dictated by Rust's
lifetime discipline. Rather, it establishes what is allowed with regards to ownership in the
presence of moves, borrows and copies. As such, our semantics can account not
only for the current borrow-checker's behavior, but also for its future
evolutions, such as Polonius~\cite{polonius}.

Our functional semantics paves the way for our functional translation. We
proceed in two steps. First, we tweak our semantics to abstract the
aliasing graph in the presence of function calls; to do so, we interpret
regions as bags of borrows and loans. Next, we follow the structure of the
program in the presence of these region abstractions, and generate a functional
translation. To overcome the key difficulty of terminating a borrow, we rely on
a technical innovation called \emph{backward functions}, which obviates the need
for prophecy variables (as in RustHorn), or \li+after+ clauses (as in Verus).

We have implemented our functional translation approach in a mixture of Rust and
OCaml, for a total of 10,000 and 14,000 lines of code, respectively (excluding
comments and whitespace). Once the pure translation is synthesized, we emit pure code for the
theorem prover of our choice: we currently support \fstar, and we have a Coq
backend in the works.

We evaluate \aeneas on a wide variety of micro-benchmarks for
feature-completeness, and verify a resizable hash table as our main
case study.
We find that the benefits of \aeneas are many. First,
with \aeneas, the verification engineer deals with the
\emph{intrinsic} difficulty of the proofs, rather than the \emph{incidental}
complexity; that is, they can focus on the essence of the proof rather than
being mired in the technicalities of memory reasoning. There are no
modifies-clause lemmas, and no separation logic framing tactics; these are,
by construction, un-necessary.
Second, our approach remains \emph{lightweight}: we do not need to design an
annotation language for Rust programs, and the properties we prove are not
constrained by the expressivity or the usability of said annotation language.
Consequently, \aeneas-translated programs can easily integrate within an
existing project, and leverage libraries or proof tactics, as opposed to
evolving in a closed world whose boundaries are set by a specific annotation
language.
Third, we are not beholden to one specific verification framework; we envision a
world where the verification engineer can simply direct \aeneas towards the
proof framework they are the most productive in.

In short, \aeneas reveals the
functional essence of Rust programs, and verifies them as such.

We start with an accessible, example-based introduction to
Rust and \aeneas (\sref{examples}), then present this paper's contributions:
\begin{itemize}
  \item an ownership-centric operational semantics for Rust (\sref{semantics}),
  \item a concept of \emph{region abstraction},
    which precisely models the
    interaction of borrows, regions and function calls; region abstractions establish
    a blueprint for our functional translation, which relies on what we dub
    \emph{backward functions} (\sref{translation}),
  \item a complete implementation, based on a Rust plugin and a subsequent OCaml
    compiler named \aeneas (\sref{implementation});
    the former extracts all the information we need from the internal Rust AST,
    while the latter implements all of the steps described in this paper,
  \item an experimental evaluation that culminates in the verification of a
    resizable hash table; to the best of our knowledge, the man-hours spent
    proving that the Rust hash table functionally behaves like a map are
    extremely modest
    relative to other
    similar efforts: \aeneas thus offers substantial gains of productivity (\sref{evaluation}).
\end{itemize}
We acknowledge the current limitations of our tool; situate \aeneas relative to
other Rust verification efforts, and conclude (\sref{future}). The
implementation of our tools, the verified code and a long version of this paper
are available online~\cite{aeneas-website,longversion,artifact}.

\section{\aeneas and its Functional Translation, by Example}
\label{sec:examples}

Before jumping into the various facets of our formalism, we keep an eye on the
prize, and immediately showcase how \aeneas translates Rust programs to pure
equivalents. In this section, and for the remainder of the paper, we use
\fstar syntax for our functional translation; it greatly resembles OCaml and
other ML languages, and as such should be familiar to the reader. A brief note
about terminology: we adopt the view of~\citet{niko-regions}, and refer to
\emph{regions}, emphasizing that a region encompasses a set of borrows and
loans at a given program point. The Rust compiler and documentation, however,
refer to \emph{lifetimes}, which conveys the idea of a syntactic bracket, and
a specific implementation technique to enforce soundness. In this paper,
whenever we talk about Rust specifically, we use ``lifetime''; whenever we emphasize
our semantic view of ownership, we use ``region''.

\myparagraph{Mutable Borrows, Functionally}
To warm up, we consider an example that, albeit small, showcases
many of Rust's
features, including its ownership mechanism. In the Rust program below,
\li+ref_incr+ increments a reference, and \li+test_incr+ acts as a
representative caller of the function.

\begin{minted}{rust}
fn ref_incr(x: &mut i32) {
    *x = *x + 1; }

fn test_incr() {
    let mut y = 0i32;
    ref_incr(&mut y);
    assert!(y == 1); }
\end{minted}

The \li+incr+ function operates by reference; that is, it receives the address of
a 32-bit signed integer \li+x+, as indicated by the \li+&+ (reference) type. In
addition, \li+incr+ is allowed to modify the contents at address \li+x+, because
the reference is of the \li+mut+ (mutable) kind, which permits memory
modification. Finally, the Rust type system enforces that mutable references
have a unique owner: the definition of \li+ref_incr+ type-checks, meaning that
the function not only \emph{guarantees} it does not duplicate ownership of
\li+x+, but also can \emph{rely} on the fact that no one else owns \li+x+.

In \li+test_incr+, we allocate a mutable value (\li+let mut+) on the stack; upon
calling \li+ref_incr+, we take a mutable reference (\li+&mut+) to \li+y+.
Statically, \li+y+ becomes unavailable as long as \li+&mut y+ is active. In Rust
parlance, \li+y+ is \emph{mutably borrowed} and its ownership has been
transfered to the mutable reference.
To type-check the call, the type-checker performs a lifetime analysis: the
\li+ref_incr+ function has type \li+(&'a mut i32) -> ()+, and the \li+&mut y+
borrow has type \li+&'b mut i32+; both \li+'a+ and \li+'b+ are lifetime variables.

For now, suffices to say that the type-checker ascertains that the lifetime
\li+'b+ of the mutable borrow satisfies the lifetime annotation \li+'a+ in the
type of the callee, and deems the call valid.
Immediately after the call, Rust \emph{terminates} the
region \li+'b+, in effect \emph{relinquishing} ownership of the mutable reference
\li+&'b mut y+ so as to make \li+y+ usable again inside \li+test_incr+.
This in turn allows the
\li+assert+ to type-check, and thus the whole program.
Undoubtedly, this is a very minimalistic program; yet, there are two properties
of interest that we may want to establish already. The obvious one: the
assertion always succeeds. More subtly, doing so requires us to prove an
additional property, namely that the addition at line 2 does not overflow.

The key insight of \aeneas is that even though the program manipulates
references and borrows, none of this is informative when it comes to reasoning
about the program. More precisely: \li+x+ and \li+y+ are uniquely owned, meaning
that there are no stray aliases through which \li+x+ or \li+y+ may be modified;
in other words, to understand what happens to \li+y+, it suffices to track what
happens to \li+&mut y+, and therefore to \li+x+.
Feeding this program to \aeneas generates the following translation, where
\li+i32_add+ is an \aeneas primitive that captures the semantics of
error-on-overflow in Rust.

\begin{minted}{fstar}
let ref_incr_fwd (x : i32): result i32 =
  match i32_add x 1 with | Fail -> Fail | Return x0 -> Return x0

let test_incr : result unit =
  match ref_incr_fwd 0 with
  | Fail -> Fail
  | Return y -> if not (y = 1) then Fail else Return ()

let _ = assert (test_incr = Return ())
\end{minted}
This program is semantically equivalent to the original Rust code, but does not
rely on the memory: we have leveraged the precise ownership discipline of Rust
to generate a functional, pure version of the program. In hindsight, the usage
of references in Rust was merely an implementation detail, which is why
\li+ref_incr_fwd+ becomes a simple (possibly-overflowing) addition. Should the
call to \li+ref_incr_fwd+ (line 5) succeed, its result is bound to \li+y+ (line
7); the \li+assert+ simply becomes a boolean test that may generate a failure
in the error monad.

For the purposes of unit-testing, \aeneas inserts an additional assertion for
\li+test_*+ functions of type $\mathsf{unit}\to\mathsf{unit}$: the prover shows instantly that our test
always succeeds. (In \fstar, we execute this assertion directly on the
normalizer, without even resorting to SMT; \aeneas produces an executable
translation, not a logical encoding.)
In the remainder of this section, we use \mintinline{fstar}+<--+, \fstar's bind
operator%
\footnote{There are issues in \fstar related to this
notation~\cite{fstar1288}, which we work around in practice, but ignore here.}
in the error monad.

\myparagraph{Returning a Mutable Borrow, and a Backward Function}
\label{sec:example:choose}
Rust programs, however, rarely admit such immediate translations. To see why,
consider the following example, where the \li+choose+ function returns a borrow,
as indicated by its return type \li+&'a mut+.

\begin{minted}{rust}
fn choose<'a, T>(b: bool, x: &'a mut T, y: &'a mut T) -> &'a mut T {
    if b { return x; } else { return y; } }

fn test_choose() {
    let mut x = 0i32; let mut y = 0i32;
    let z = choose(true, &mut x, &mut y);
    *z = *z + 1;
    assert!(*z == 1);
    assert!(x == 1); assert!(y == 0); }
\end{minted}
The \li+choose+ function is polymorphic over type \li+T+ and lifetime \li+'a+; the
lifetime annotation captures the expectation that both \li+x+ and \li+y+ be in the
same region.
At call site, \li+x+ and \li+y+ are borrowed (line 6): they become unusable,
and give birth to two intermediary values \li+&mut x+ and \li+&mut y+ of type
\li+&'a mut i32+. The value returned by \li+choose+ also lives in region
\li+'a+, i.e., \li+z+ also has type \li+&'a mut i32+.
The usage of \li+z+ (lines 7-8) is valid because the region \li+'a+ still
exists; the Rust type-checker infers that region \li+'a+ ought to be terminated
after line 8, which ends the borrows and therefore allows the caller to regain
full ownership of \li+x+ and \li+y+, so that the \li+assert+s at line 9 are
well-formed.

At first glance, it appears we can translate \li+choose+ to an obvious
conditional.
But if we reason about the semantics of \li+choose+ from the
caller's perspective, it turns out that the intuitive translation is not
sufficient to capture what happens, e.g., to \li+x+ and \li+y+ at lines 9.
At call site, \li+choose+ is an opaque, separate function, meaning the caller
cannot reason about its precise definition -- all that is
available is the function type. This type, however, contains precise region
information. When performing the function call, the ownership of \li+x+ and
\li+y+ is transferred to region \li+'a+ in exchange for \li+z+; symmetrically,
when the lifetime
 \li+'a+ terminates, \li+z+ is relinquished to region
\li+'a+ in exchange for regaining ownership of \li+x+ and \li+y+. The former
operation flows \emph{forward}; the latter flows \emph{backward}. Using a
separation-logic oriented analogy: borrows and regions encode a magic wand that
is introduced in a function call and eliminated when the corresponding region
terminates.

Our point is: both function call and region termination are semantically
meaningful. In our earlier example, the \li+ref_incr+ function returned a unit,
meaning that \aeneas only emitted a \emph{forward} function (hence the
\li+_fwd+ suffix) to translate the function call.
With the \li+choose+ example,
\aeneas emits both a \emph{forward} and a \emph{backward} function, used for the
function call and the end of the \li+'a+ region, respectively.
\begin{minted}{fstar}
let choose_fwd (t : Type) (b : bool) (x : t) (y : t) : result t =
  if b then Return x else Return y

let choose_back (t : Type) (b : bool) (x : t) (y : t) (ret : t) : result (t & t) =
  if b then Return (ret, y) else Return (x, ret)

let test_choose_fwd : result unit =
  i <-- choose_fwd i32 true 0 0;
  z <-- i32_add i 1;
  massert (z = 1); (* monadic assert *)
  (x0, y0) <-- choose_back i32 true 0 0 z;
  massert (x0 = 1);
  massert (y0 = 0);
  Return ()

let _ = assert (test_choose_fwd = Return ())
\end{minted}
The call to \li+choose+ becomes a call to the forward function \li+choose_fwd+
(line 8); we bind the result of the addition (provided no overflow occurs) to
\li+z+ (line 9); then, per the rules of Rust's type-checker, region
\li+'a+ terminates which compels us to call the backward function \li+choose_back+.
The intuitive effect of calling \li+choose_back+ is as follows: we relinquish
\li+z+, which was in region \li+'a+; doing so, we propagate any updates
that may have been performed through \li+z+ onto the
variables whose ownership was
transferred to \li+'a+ in the first place, namely \li+x+ and \li+y+.
This bidirectional approach is akin to lenses~\cite{bohannon2008boomerang}, except we propagate
the output back to possibly-many inputs; in this case, \li+z+ is a view
over either \li+x+ or \li+y+, and the backward function reflects the update to
\li+z+ onto the original variables.
Thus, both variables are re-bound (line 11),
before chaining the two asserts (lines 12 and 13).

From the caller's perspective, the computational content of \li+choose+ is
unknown; but the signature of \li+choose+ reveals the effect it may have onto its
inputs \li+x+ and \li+y+, which in turns allows us to derive the type of the
backward and forward functions from the signature of \li+choose+ itself.
The result is a modular, functional
translation that does not rely on any sort of cross-function inlining or
whole-program analysis.
To synthesize \li+choose_back+, it suffices to invert the direction
of assignments; in one case, \li+z+ flows to \li+x+ and \li+y+ remains
unchanged; the other case is symmetrical.

\myparagraph{Recursion and Data Structures}
It might not be immediately obvious that this translation technique scales up
beyond toy examples; to conclude this section, we crank up the complexity and
show how \aeneas can handle a wide variety of idioms while still delivering on the
original promise of a lightweight functional translation.
Our final example is
 \li+list_nth+, which allows taking a mutable reference to the $n$-th
element of a list, mutating it, and regaining ownership of the list.
\begin{minted}{rust}
enum List<T> { Cons(T, Box<List<T>>), Nil }

fn list_nth_mut<'a, T>(l: &'a mut List<T>, i: u32) -> &'a mut T {
    match l {
        Nil => { panic!() }
        Cons(x, tl) => { if i == 0 { x } else { list_nth_mut(tl, i - 1) } } } }

fn sum(l: & List<i32>) -> i32 {
    match l {
        Nil => { return 0; }
        Cons (x, tl) => { return *x + sum(tl); } } }

fn test_nth() {
    let mut l = Cons (1, Box::new(Cons (2, Box::new(Cons (3, Box::new(Nil))))));
    let x = list_nth_mut(&mut l, 2);
    *x = *x + 1;
    assert!(sum(&l) == 7); }
\end{minted}
This example relies on several new concepts. Parametric data type declarations
(line 1) resemble those in any functional programming language such as OCaml
or SML. The \li+Box+ type denotes a heap-allocated, uniquely-owned piece of
data. Without the \li+Box+ indirection, \li+List+ would describe a type of
infinite size and would be rejected. Immutable borrows (line 8) do not sport a
\li+mut+ keyword; they do not permit mutation, but the programmer may create
infinitely many of them. Only when all shared borrows have been relinquished
does full ownership return to the borrowed value.
\ifshort
The complete translation is in our long version~\cite{longversion}; the salient part comes
out of \aeneas as follows:
\begin{minted}[mathescape=true,escapeinside=!!]{fstar}
type list_t (t : Type) = | ListCons : t -> list_t t -> list_t t | ListNil : list_t t

let rec list_nth_mut_back (t : Type) (l : list_t t) (i : u32) (ret : t) : result (list_t t) =
  match l with
  | ListNil ->       Fail
  | ListCons x tl -> match i with
                     | 0 -> Return (ListCons ret tl)
                     | _ -> i0 <-- u32_sub i 1;
                            l0 <-- list_nth_mut_back t tl i0 ret;
                            Return (ListCons x l0)

let test_nth_fwd : result unit =
  let l2 = ListCons (1, ListCons (2, ListConst (3, ListNil))) in
  i <-- list_nth_mut_fwd i32 l2 2; !\label{line:nth0}!
  x <-- i32_add i 1; !\label{line:nth1}!
  l2 <-- list_nth_mut_back i32 l2 2 x; !\label{line:nth2}!
  i0 <-- sum_fwd l2;
  massert (i0 = 7)
\end{minted}
\fi
\iflong
The complete translation is as follows:
\begin{minted}[mathescape=true,escapeinside=!!]{fstar}
type list_t (t : Type) =
| ListCons : t -> list_t t -> list_t t
| ListNil : list_t t

let rec list_nth_mut_fwd (t : Type) (l : list_t t) (i : u32) : result t =
  begin match l with
  | ListCons x tl ->
    begin match i with
    | 0 -> Return x
    | _ ->
      i0 <-- u32_sub i 1;
      list_nth_mut_fwd t tl i0
    end
  | ListNil -> Fail
  end

let rec list_nth_mut_back (t : Type) (l : list_t t) (i : u32) (ret : t) : result (list_t t) =
  begin match l with
  | ListCons x tl ->
    begin match i with
    | 0 -> Return (ListCons ret tl)
    | _ ->
      i0 <-- u32_sub i 1;
      l0 <-- list_nth_mut_back t tl i0 ret;
      Return (ListCons x l0)
    end
  | ListNil -> Fail
  end

let rec sum_fwd (l : list_t i32) : result i32 =
  begin match l with
  | ListCons x tl ->
    i <-- sum_fwd tl;
    i32_add x i
    end
  | ListNil -> Return 0
  end

let test_nth_fwd : result unit =
  let l2 = ListCons (1, ListCons (2, ListConst (3, ListNil))) in
  i <-- list_nth_mut_fwd i32 l2 2; !\label{line:nth0}!
  x <-- i32_add i 1; !\label{line:nth1}!
  l2 <-- list_nth_mut_back i32 l2 2 x; !\label{line:nth2}!
  i0 <-- sum_fwd l2;
  massert (not (i0 = 7))

let _ = assert (test_nth_fwd = Return ())
\end{minted}
\fi
We first focus on the caller's point of view.
Continuing with the lens analogy, we focus on (or ``get'') the $n$-th element of the list via
a call to \li+list_nth_mut_fwd+ (line \ref{line:nth0}); modify the element (line
\ref{line:nth1}); then
close (or ``put'' back) the lens, and propagate the modification back to the list via a call to
\li+list_nth_mut_back+ (line \ref{line:nth2}).
The \li+list_nth_mut_back+ function is of particular interest. The function
follows the same control-flow as the forward function.
However, the key part happens when returning from the \li+Cons+ case: our
functional translation performs a \emph{semantic} region analysis, from which it follows that a
recursive call to the backward function is needed, along with the construction
of a new \li+Cons+ cell.

\section{An Ownership-Centric Semantics for Rust}
\label{sec:semantics}

Before explaining the functional translation above,
we must first define our
input language and its operational semantics. We now present a series of short
Rust snippets, and show in comments how our execution environments model the
effect of each statement.

\myparagraph{Mutable borrows}
After line 1, $x$ points to $0$, which we write $x \mapsto 0$. At line 2, \li+px+ \emph{mutably} borrows \li+x+. As we
mentioned earlier, a mutable borrow grants exclusive ownership of a value, and
renders the borrowed value unusable for the duration of the borrow. We reflect this
fact in our execution environment as follows: \li+x+ is marked as
``loaned-out'', in a \underline{m}utable fashion, and \li+px+ is known to be a
\underline{m}utable borrow. Furthermore, ownership of the borrowed value now
rests with \li+px+, so the value within the mutable borrow is 0. Finally, we
need to record that \li+px+ is a borrow \emph{of \li+x+}: we issue a fresh loan
identifier $\ell$ that ties \li+x+ and \li+px+ together.
The same operation is repeated at line 3. Value 0 is now held by \li+ppx+, and
\li+px+, too, becomes ``loaned out''.
\begin{minted}[mathescape=true]{rust}
let mut x = 0;       // $x \mapsto 0$
let mut px = &mut x; // $x \mapsto \emloan\ell,\quad px \mapsto \emborrow\ell0$
let ppx = &mut px;   // $x \mapsto \emloan\ell,\quad px \mapsto \emloan\ell',\qquad ppx \mapsto \emborrow{\ell'}{(\emborrow\ell0)}$
\end{minted}
Our environments thus precisely track ownership; doing so, they capture the aliasing
graph in an exact fashion. Another point about our style:
this representation allows us to adopt a \emph{focused} view of the borrowed
value (e.g. 0), solely
through its owner (e.g. \li+ppx+), without worrying about following indirections
to other variables.
We believe this
approach is unique to our semantics; it has, in our experience, greatly
simplified our reasoning and in particular the functional translation
(\sref{translation}).

We remark that our style
departs from Stacked Borrows~\cite{jung2019stacked}, where the modified value
remains with \li+x+. We also note that our formalism cannot account for unsafe
blocks; allowing unfettered aliasing would lead to potential cycles, which we
cannot represent. This is an intentional design choice for us: we circumbscribe
the problem space in order to achieve an intuitive, natural semantics and a
lightweight functional translation. \aeneas shines on non-unsafe Rust programs,
and can be complemented by more advanced tools such as RustBelt for unsafe
parts.

\myparagraph{Shared borrows}
Shared borrows behave more like traditional pointers. Multiple shared borrows
may be created for the same value; each of them grants read-only access to the
underlying value. Regaining full ownership of the borrowed value requires
terminating all of the borrows. In the example
below, the value $(0, 1)$ is borrowed in a shared fashion, at line 2. This time,
the value remains with \li+x+; but taking an immutable reference to
\li+x+ still requires book-keeping. We issue a new loan $\ell$, and record that
\li+px1+ is now a shared borrow associated to loan $\ell$; to understand which
value \li+px1+ points to, we simply look up in the environment who is the owner
of $\ell$, and read the associated value.
Repeated shared borrows are permitted: at line 3, we issue a new loan $\ell'$
which augments the loan-set of \li+x+.
At line 4, we anticipate on our internal syntax, where moves and copies are
explicit, and copy the first component of \li+x+. Values that are loaned
immutably, like \li+x+, can still be read; in the resulting environment, \li+y+ points to a copy of
the first component, and bears no relationship whatsoever to \li+x+.
Finally, at line 5, we \emph{reborrow} (through \li+px1+) the first component of
the pair only. First, to dereference \li+px1+, we perform a lookup and find that
\li+x+ owns $\ell$. Then, we perform book-keeping and update the value loaned by
\li+x+, so as to reflect that its first component has been loaned
out.
\begin{minted}[mathescape]{rust}
let x = (0, 1);    // $x \mapsto (0, 1)$
let px1 = &x;      // $x \mapsto \esloan\ell(0,1),$  $\qquad px1 \mapsto \esborrow\ell$
let px2 = &x;      // $x \mapsto \esloan{\ell\cup\ell'}(0, 1),\quad px1 \mapsto \esborrow\ell,\quad px2 \mapsto \esborrow\ell'$
let y = copy x.0;  // $x \mapsto \esloan{\ell\cup\ell'}(0, 1),\quad px1 \mapsto \esborrow\ell,\quad px2 \mapsto \esborrow\ell',\quad y \mapsto 0$
let z = &(*px1.0); // $x \mapsto \esloan{\ell\cup\ell'}((\esloan{\ell''} 0), 1),\ px1 \mapsto\ldots,\ px2 \mapsto \ldots,\ y \mapsto 0,\ z \mapsto \esborrow\ell''$
\end{minted}
\iflong
In our presentation, shared borrows behave like pointers, and every one of them
is statically accounted for via the set of loans attached to the borrowed value.
The reader might find this design choice surprising: indeed, in Rust, shared
borrows behave like immutable values and we ought to be able to treat them as
such. Recall, however, that one of our key design goals is to give a
\emph{semantic} explanation of borrows; as such, our precise tracking of
shared borrows allows us to know precisely when all aliases
have been relinquished, and full ownership of the borrowed value has been
regained. This allows us to justify why, in the example below, the update to
\li+x+ is sound; without our exact alias tracking, we would have to trust the
borrow checker, something we explicitly do not want to do.
\begin{minted}[mathescape]{rust}
let mut x = 0;  // $x \mapsto 0$
let px1 = &x;   // $x \mapsto \esloan\ell0,$  $\qquad px1 \mapsto \esborrow\ell$
let px2 = &x;   // $x \mapsto \esloan{\ell\cup\ell'}0,\quad px1 \mapsto \esborrow\ell,\quad px2 \mapsto \esborrow\ell'$
                // $x \mapsto 0,\hspace{6.8em} px1 \mapsto \bot,\quad px2 \mapsto \bot$
x = 1;          // $x \mapsto 1$
\end{minted}
\fi
Finally, we reiterate our remark that our formalism allows keeping track of the aliasing
graph in a precise fashion; the discipline of Rust bans cycles, meaning that the
aliasing graph is always a tree. This style of representation resembles
Mezzo~\cite{balabonski2016design}, where loan identifiers are akin to singleton
types, and entries in the environment are akin to permissions.

\myparagraph{Rewriting an Old Value, a.k.a. Reborrowing}
We now consider a particularly twisted example accepted by the Rust compiler. While the
complexity seems at first gratuitous, it turns out that the pattern of borrowing
a dereference (i.e., \li+&mut (*px)+) is particularly common in Rust. The reason
is subtle: in the post-desugaring MIR internal Rust representation, moves and
copies are explicit, meaning function calls of the form \li+f(move px)+ abound.
Such function calls \emph{consume} their argument, and render the
user-declared reference \li+px+ unusable past the function call.
To offer a better user experience, Rust automatically ``reborrows'' the
contents pointed to by \li+px+, and rewrites the call into \li+f(move (&mut (*px)))+ at desugaring-time.
Thus, only the intermediary value is ``lost'' to the function call;
relying on its lifetime analysis, the Rust compiler concludes that the
user-declared reference \li+px+ remains valid past the function call, hence
making the programmer's life easier.

Another common pattern is to directly mutate a borrow, i.e. assign a fresh
borrow into a variable \li+x+ of type \li+&mut t+ that was \emph{itself}
declared as \li+let mut+.
Capturing the semantics of such an update must be done with great care, in order
to preserve precise aliasing information.

We propose an example that combines both patterns; the fact that we make
\li+px+ reborrow itself is what makes the example ``twisted''. Rust accepts this
program; we now explain with our semantics \emph{why} it is sound.
In the example below, after line
2, the environment offers no surprises. Justifying the write at line 3 requires
care. We borrow $*px$, which modifies $px$ to point to $\emborrow{\ell'}0$,
and returns $\emloan{\ell'}$; the value about to be overwritten is stored in a
fresh variable $px_\mathsf{old}$, and $\emloan\ell'$
gets written to $px$.
\begin{minted}[mathescape]{rust}
let mut x = 0;       // $x \mapsto 0$
let mut px = &mut x; // $x \mapsto \emloan\ell,\quad px \mapsto \emborrow\ell0$
px = &mut (*px);     // $x \mapsto \emloan\ell,\quad px_{\mathsf{old}} \mapsto \emborrow\ell{(\emloan\ell')},\quad px \mapsto \emborrow\ell'0$
                     // $x \mapsto \emloan\ell,\quad px_{\mathsf{old}} \mapsto \emborrow\ell0,\hspace{4.8em} px \mapsto\bot$
assert!(x==0);       // $x \mapsto 0,\hspace{3.7em} px_{\mathsf{old}} \mapsto \bot,\hspace{9.2em} px \mapsto \bot$
\end{minted}
Saving the old value is crucial for line 4.
For the assertion, we need to regain full ownership of $x$. To do so,
we first terminate $\ell'$.
This \emph{reorganizes} the environment, with two consequences.
First, $px$ becomes unusable, which we write $px \mapsto
\bot$. Second, $px_\mathsf{old}$, which we had judiciously kept in the
environment, becomes $\emborrow\ell0$. We reorganize the environment again, to
terminate $\ell$; the effect is similar, and results in $x \mapsto
0$, i.e. full ownership of $x$.
This example illustrates a key characteristic of our approach, which is that we
reorganize borrows in a lazy fashion, and don't terminate a borrow until we
need to get the borrowed value back.

\iflong
\myparagraph{An Illegal Borrow}
We offer an final example which leverages our reorganization rules;
furthermore, the example illustrates how our semantics reaches the same
conclusion as \li+rustc+, though by different means, on a borrowing error.
At line 2, a borrow is introduced, which results in a fresh loan $\ell$.
To make progress, we
terminate borrow $\ell$ at line 3.
Line 4 then type-checks, with a fresh borrow $\ell'$. Then, we error
out at line 5: $px_1$ has been terminated, and we cannot dereference $\bot$.

The Rust compiler proceeds differently, and implements an analysis which
requires computing borrow constraints for an entire function body.
The compiler notices that lifetime $\ell$ must go on until line 6 (because of
the assert), which prevents a new borrow from being issued at line 4. Rust
thus ascribes the error to an earlier location than we do, that is, line 4.
We remark that the Rust behavior is semantically equivalent to ours; however, our
lazy approach which terminates borrows only as needed has the advantage that
evaluation can proceed in a purely forward fashion, without requiring a
non-local analysis. This on-demand approach is similar to Stacked
Borrows.

\begin{minted}[mathescape]{rust}
let mut x = 0;    // $x \mapsto 0$
let px1 = &mut x; // $x \mapsto \emloan\ell,\quad px_1 \mapsto \emborrow\ell0$
                  // $x \mapsto 0,\quad px_1 \mapsto \bot$
let px2 = &mut x; // aeneas: $x \mapsto \emloan\ell',\quad px_1 \mapsto \bot, \quad px_2 \mapsto \emborrow{\ell'}0$
                  // rustc: error: cannot borrow `px1` as mutable more than once at a time
assert!(*px1 == 0); // aeneas: error, attempt to deference unusable variable px1
\end{minted}
\fi

\subsection{The Low-Level Borrow Calculus}
\label{sec:llbc}

\newcommand*\ruleline[1]{\par\noindent\raisebox{.5ex}{\makebox[\linewidth]{\hrulefill\hspace{1ex}\raisebox{-.5ex}{#1}\hspace{1ex}\hrulefill}}}

\begin{figure}
    \smaller
    \arraycolsep=1pt %
    \centering

    \ifshort
    \ruleline{\sffamily\textbf{Syntax}} %
    \vspace{-2ex} %
    \fi
    \[
    \ifshort
    \begin{array}{ll} %
    \fi
    \begin{array}[t]{llll}
      \tau & ::= & & \text{type} \\
        &&
        \kw{bool}\mid\kw{uint32}\mid\kw{int32}\iflong\mid\ldots\fi & \text{base types} \\
        && \tmbrw\rho\tau & \text{mutable borrow} \\
        && \tbrw\rho\tau & \text{immutable \iflong(shared) \fi borrow} \\
        && T\; \vec \tau & \text{type application} \\
        && \alpha, \beta, \ldots & \text{type variables} \\
        && (\vec \tau) & \text{tuple \iflong($\kw{len}(\vec \tau) > 1$) \fi or unit \iflong($\kw{len}(\vec \tau) = 0$)\fi}
      \\[1ex]

      T & ::= & & \text{type \iflong constructor\fi application} \\
      && t& \text{user-defined data type\ifshort\quad\fi} \\
      && \kw{Box}& \text{boxed type}
      \\[1ex]

      s & ::= & & \text{statement} \\
      && \emptyset & \text{empty statement (nil)} \\
      && \eseq s s & \text{sequence (cons)} \\
      && \eassign p \krv & \text{assignment} \\
      && \eassign p {f\,\vec\kop} & \text{function call} \\
      && \eite \kop s s & \text{conditional} \\
      && \ematch p {\overrightarrow{ C \to s}} & \text{\iflong data type\fi case analysis} \\
      && \ereturn & \text{function exit} \\
      && \epanic & \text{unrecoverable error} \\
      && \ldots & \text{loops, etc.}
      \\[1ex]

      rv & ::= & & \text{assignable ``r'' values} \\
      && \kop & \text{operand} \\
      && \embrw p & \text{mutable borrow} \\
      && \ebrw p & \text{immutable \iflong(shared) \fi borrow} \\
      && !\kop \mid \kop + \kop \mid \kop - \kop \iflong\mid \ldots\fi & \text{operators}
      \\[1ex]

    \ifshort
    \end{array} & %
    \begin{array}[t]{llll} %
    \fi

      \kop & ::= & & \text{operand} \\
      && \emove p & \text{ownership transfer} \\
      && \ecopy p & \text{scalar copy} \\
      && \ktrue \mid \kfalse \mid n_{\mathsf{i32}} \mid n_{\mathsf{u32}} \iflong\mid \ldots\fi & \text{constants} \\
      && C [\vec f = \vec{\kop}] & \text{data type constructor} \\
      && (\vec{op}) & \text{tuple \iflong($\kw{len}(\vec{op}) > 1$)\fi or unit \iflong($\kw{len}(\vec{op}) = 0$)\fi}
      \\[1ex]

      x & & & \text{variable}
      \\[1ex]

      p & ::= & P[x] & \text{place}
      \\[1ex]

      P & ::= & & \text{path} \\
      && [.] & \text{base case} \\
      && \ederefs P & \text{deref shared borrow} \\
      && \ederefm P & \text{deref mutable borrow} \\
      && \ederefb P & \text{deref box} \\
      && P.f & \text{field selection} \\
      && P.n & \text{field selection (tuple)}
      \\[1ex]

      D & ::= & & \text{top-level declaration} \\
      &&
      \iflong
      \mathsf{fn}\,f\,\langle\vec\rho\rangle
        \ (\vec x_\mathsf{arg}: \vec\tau)
        \ (\vec x_\mathsf{local}: \vec \tau)
        \ (x_\mathsf{ret}: \tau)
        = s\quad
        \fi
        \ifshort
      \begin{array}l
      \mathsf{fn}\,f\,\langle\vec\rho\rangle \\
        \ (\vec x_\mathsf{arg}: \vec\tau)\\
        \ (\vec x_\mathsf{local}: \vec \tau)\\
        \ (x_\mathsf{ret}: \vec \tau)\\
        = s
      \end{array}
      \fi
        & \text{function declaration} \\
      && \mathsf{type}\,t\;\vec\alpha = C [\vec f: \vec\tau] \mid \ldots & \text{data type declaration}
    \end{array}
    \ifshort
    \end{array} %
    \fi
    \]

    \iflong
  \caption{The Low-Level Borrow Calculus: Syntax} %
  \label{fig:syntax} %
\end{figure} %

\begin{figure} %
    \smaller %
    \centering %
  \fi
    \ifshort
      \ruleline{\sffamily\textbf{Reduction}}
      \centering
      \arraycolsep=4pt
    \fi
    \[
    \ifshort
    \begin{array}{ll}
    \fi
    \begin{array}[t]{llll}
      v & ::= & & \text{value} \\
        && \ktrue \mid \kfalse \mid n_{\mathsf{i32}} \mid n_{\mathsf{u32}} \iflong\mid \ldots\quad\fi & \text{booleans and integers} \\
        && \emborrow \ell v & \text{mutable borrow} \\
        && \esborrow \ell & \text{shared borrow} \\
        && \eiborrow \ell & \text{reserved borrow} \\
        && \emloan \ell & \text{mutable loan} \\
        && \esloan {\vec\ell} v & \text{shared loan} \\
        && \bot & \text{inaccessible value} \\
        && C\,[ \vec f = \vec v ] & \text{data type constructor} \\
        && (\vec v) & \text{tuple constructor \iflong($\kw{len}(\vec v) > 1$) \fi or unit \iflong($\kw{len}(\vec v) = 0$)\fi}
      \\[1ex]

      r & ::= & & \text{result} \\
        && \epanic & \text{unrecoverable error state} \\
        && \ereturn\,v & \text{successful, possibly-early exit} \\
        && \ldots & \text{loop states, etc.}
      \\[1ex]

    \ifshort
    \end{array}
    &
    \begin{array}[t]{llll}
    \fi

      \ell & & & \text{loan identifier}
      \\[1ex]

      \Omega & ::= & & \text{evaluation context} \\
        && \emptyset & \text{empty} \\
        && x \mapsto v, \Omega & \text{new mapping} \\
    \end{array}
    \ifshort
    \end{array}
    \fi
    \]
    \ifshort
  \caption{The Low-Level Borrow Calculus: Syntax, Reduction Environments, Values}
    \fi
    \iflong
  \caption{The Low-Level Borrow Calculus: Reduction Environments, Values}
  \fi
  \label{fig:values}
    \ifshort
  \label{fig:syntax}
    \fi
\end{figure}

We now formally introduce and define our semantics of Rust programs. We start
with the Low-Level Borrow Calculus (``LLBC'', \fref{syntax}), our source
language. LLBC is in large part inspired by MIR, Rust's post-desugaring internal
representation, notably: all local variables $\vec x_\mathsf{local}$ are bound at the beginning of the
function declaration; returning a value from a function amounts to writing into
the special variable $x_\mathsf{ret}$, followed by $\ereturn$; all subexpressions have been
named so as to fit within MIR's statement/r-value/operand categories; and all
variables within expressions have been desugared to either a move, a copy or a
borrow. %
However, and unlike MIR, LLBC retains some
high-level constructs: control-flow remains structured (LLBC statements are thus the
fusion of MIR's statements and terminators); and case analysis on data types
is exposed via a limited form of (complete) pattern matching, as opposed to a low-level integer switch on the
tag. We remark that data types may match on a \emph{path} only; this merely imposes that the
scrutinee be let-bound before examining it, something that MIR does internally.
Finally, we use pure expressions to allocate data types, rather than
the progressive (mutable) initialization pattern used by MIR; and we see structures as
data types equipped with a single constructor, for conciseness.
Staying close to MIR is a design choice in line with other Rust-related
works~\cite{jung2019stacked}; it allows for fewer, simpler
rules, and a more
precise description of what happens from the point of view of ownership.

At the heart of LLBC is a notion of \emph{place}, i.e. the combination of a base variable
(e.g. \li+x+) and a series of field offsets and indirections known as a
\emph{path} (e.g. \li+*_.f+). A place
is akin to the notion of ``lvalue'' in, e.g., C. Assigning or returning from a
function can only be done into a \emph{place}. The grammar of rvalues and
operands (Rust's limited form of expression) is very explicit, in that
every use of a variable is performed through a copy, a move, or a borrow of a place.

\subsection{A Structured Memory Model}

\begin{figure}
  \centering
  \smaller
  \begin{mathpar}
    \inferrule[Read]{
      p = P[x] \\
      x \mapsto v_x \in \Omega \\
      \Omega \vdash P(v_x) \Rightarrow v
    }{
      \Omega(p) \Rightarrow v
    }

    \inferrule[R-Box]{
      \Omega \vdash P(v_P) \Rightarrow v
    }{
      \Omega \vdash (\ederefb P)(\ebox v_P) \Rightarrow v
    }

    \inferrule[R-Mut-Borrow]{
      \Omega \vdash P(v_P) \Rightarrow v
    }{
      \Omega \vdash (\ederefm P)(\emborrow \ell v_P) \Rightarrow v
    }

    \inferrule[R-Shared-Borrow]{
      \esloan{\ell \cup \_}{v_P} \in \Omega \\
      \Omega \vdash P(v_P) \Rightarrow v
    }{
      \Omega \vdash (\ederefs P)(\esborrow \ell) \Rightarrow v
    }

    \inferrule[R-Field]{
      \Omega \vdash P(v_P) \Rightarrow v
    }{
      \Omega \vdash (P.f)(C\,[ \overrightarrow{ f = v_P } ]) \Rightarrow v
    }

    \inferrule[R-Base]{ }{
      \Omega \vdash [.](v) \Rightarrow v
    }

    \inferrule[R-Shared-Loan]{
      P \neq [.] \\
      \Omega \vdash P(v_P) \Rightarrow v
    }{
      \Omega \vdash P(\esloan \_ v_P) \Rightarrow v
    }

    \inferrule[Write]{
      x \mapsto v_x \in \Omega \\
      p = P[x] \\\\
      \Omega \vdash P(v_x) \leftarrow v \Rightarrow v'_x \\
      \Omega' = \Omega[x \mapsto v'_x]
    }{
      \Omega(p) \leftarrow v \Rightarrow \Omega'
    }

    \inferrule[W-Box]{
      \Omega \vdash P(v_P) \leftarrow v \Rightarrow v'_P
    }{
      \Omega \vdash (\ederefb P)(\ebox {v_P}) \leftarrow v \Rightarrow \ebox {v'_P}
    }

    \inferrule[W-Mut-Borrow]{
      \Omega \vdash P(v_P) \leftarrow v \Rightarrow v'_P
    }{
      \Omega \vdash (\ederefm P)(\emborrow \ell {v_P}) \leftarrow v \Rightarrow \emborrow \ell {v'_P}
    }

    \inferrule[W-Base]{
    }{
    \Omega \vdash [.](v_P) \leftarrow v \Rightarrow v
    }\iflong

    \inferrule[W-Field]{
    f = f_i \\
    \Omega \vdash P(v_{P,i}) \leftarrow v \Rightarrow v'_{P,i} \\
    v'_{P,j} = v_{P,j} \text{ for } j\neq i
    }{
    \Omega \vdash (P.f)(C[\overrightarrow{f = v_P}]) \leftarrow v \Rightarrow
      C[\overrightarrow{f = v'_P}]
    }\fi
  \end{mathpar}
  \caption{Reading From and Writing To Our Structured Memory Model}
  \label{fig:proj}
\end{figure}

Rust marries high-level concepts, such as ownership, a strong notion of value,
and data types, with low-level concepts such as moves and copies, paths through
a base address, and modifications at depth throughout the store.
We propose a semantics that operates exclusively in terms of
values (that is, no memory addresses), yet still permits fine-grained memory
mutations as allowed by Rust.

\fref{values} presents our environments, which we write $\Omega$, and our
values, which we write $v$.
We do not distinguish between store and
environment, and use the two terms interchangeably.
The store maps variable names $x$ to values $v$: we have no notion of arbitrary memory
addresses, or pointer arithmetic. Our values $v$ are carefully crafted to model
the semantics of borrows and ownership tracking in Rust; several of them already
appeared in our earlier examples.

The combination of places, environments and values allows us to define reads and
writes already. Reads and writes are defined in terms of our \emph{structured}
memory model: we do not have any notion of memory address, but \emph{do} have a
notion of path combined with a base ``address'' (variable) $x$, that is, a
place; this permits reads and writes, at depth, through references.
We present selected rules for reading (\textsc{R-*}) and writing (\textsc{W-*}) in
\fref{proj}.

For \emph{reading}, we write $\Omega(p) \Rightarrow v$, meaning reading from
$\Omega$ at place $p$ produces $v$. Doing so requires looking up the ``base
pointer'' (variable) $x$ found in $p$ (\Rule{Read}), then deferring to an
auxiliary judgment of the form $\Omega \vdash P(v_x) \Rightarrow v$, meaning
following path $P$ into $v_x$ produces value $v$.
We can follow
path $P$ as long as the value $v_x$ is of the right shape (\Rule{R-Mut-Borrow}, \Rule{R-Field}).
Reading from a mutable borrow
requires no additional operation, since the mutable borrow uniquely owns the
value it points to (\Rule{R-Mut-Borrow}).
Conversely, reading from a shared borrow requires looking up
the owner of the loan to find the value being pointed to (\Rule{R-Shared-Borrow}).

Rule \Rule{R-Base} is our base case: if we have reached the end of the path
$P$, we simply return the value found there.
One subtlety occurs in the case of shared \emph{loans}.
Rule \Rule{R-Shared-Loan} permits reading from a value that is currently immutably
borrowed, which is allowed in Rust. However, we only do so if necessary; that is,
if we must follow further indirections in $P$ (i.e., $P \neq [.]$). If there are no further indirections (i.e., $P = [.]$), \Rule{R-Base} kicks in and
returns a value of the form $\kw{loan}^s$. This is intentional, and will prove useful for
\Rule{E-Shared-Or-Reserved-Borrow}, as we shall see shortly.

For \emph{writing}, we write $\Omega(p) \leftarrow v \Rightarrow \Omega'$,
meaning assigning value $v$ into $\Omega$ at place $p$ produces an updated
environment $\Omega'$ (\Rule{Write}). As before, we follow the
structure of $v_x$ (\Rule{Write}), and defer to an auxiliary judgment of the
form $\Omega \vdash P(v_x) \leftarrow v \Rightarrow v'_x$, which from $v_x$
computes an updated value $v'_x$ where only the subexpression selected by $P$ is
updated with $v$.
We update $x$'s entry in the environment to map to the new value
$v'_x$, denoted $\Omega[x\mapsto v'_x]$ (\Rule{Write}).
As before, the shape of $P$ determines which rule applies: we may only write through a mutable borrow
(\Rule{W-Mut-Borrow}) or a box (\Rule{W-Box}). We eventually apply
\Rule{W-Base}. We elide the remaining
rules (tuples, fields, etc.).

\subsection{Semantics of Ownership and Borrows}
\label{sec:rules}

\begin{figure}
  \centering
  \smaller
  \begin{mathpar}
    \inferrule[E-Mut-Borrow]{
      \Omega(p) \Rightarrow v \\
      \{\bot, \kw{loan}, \kw{borrow}^r\} \not\in v \\\\
      \ederefs\relax \not\in p \\
      \ell \text{ fresh} \\
      \Omega(p) \leftarrow \emloan\ell \Rightarrow \Omega'
    }{
      \Omega \vdash \embrw p \rightsquigarrow \emborrow \ell v \dashv \Omega'
    }

    \inferrule[E-Shared-Or-Reserved-Borrow]{
      \Omega(p) \Rightarrow v \\
      \{ \bot, \emloan\relax, \kw{borrow}^r\}\not\in v \\\\
      \ell \text{ fresh} \\
      \Omega' = \Omega[p \mapsto v'] \\\\
      v' = \begin{cases}
        \esloan {\vec \ell \cup \ell} v'' & \text { if } v = \esloan{\vec\ell} v'' \cr
        \esloan {\ell} v & \text { otherwise }
      \end{cases}
    }{
      \Omega \vdash \ebrw p \rightsquigarrow \kw{borrow}^{r,s} \ell \dashv \Omega'
    }

    \inferrule[E-Move]{
      \Omega(p) \Rightarrow v \\
      \{\bot, \kw{loan}, \kw{borrow}^r\}\not\in v \\\\
      \{\ederefm\relax, \ederefs\relax\}\not\in p \\
      \Omega(p) \leftarrow \bot \Rightarrow \Omega'
    }{
      \Omega \vdash \emove p \rightsquigarrow v \dashv \Omega'
    }

    \inferrule[E-Copy]{
      \Omega(p) \Rightarrow v \\
      \{\bot, \emloan\relax, \kw{borrow}^{m,r}\} \not\in v \\\\
      \Omega \vdash \ecopy v \Rightarrow v' \dashv \Omega'
    }{
      \Omega \vdash \ecopy p \rightsquigarrow v' \dashv \Omega'
    }\iflong

    \inferrule[E-Constructor]{
    \Omega_i \vdash op_i \rightsquigarrow v_i \dashv \Omega_{i+1}
    }{
    \Omega_0 \vdash C[\overrightarrow{f = \kop}] \rightsquigarrow
    C[\overrightarrow{f = v}] \dashv \Omega_n
    }

    \inferrule[E-Return]{
    \Omega_i \vdash \eassign{x_{\mathsf{local},i}}\bot \rightsquigarrow () \dashv \Omega_{i+1} \\\\
    \Omega_n (x_\mathsf{ret}) \Rightarrow v \\
    \{\bot, \kw{loan}\relax, \kw{borrow}^{r}\} \not\in v
    }{
      \Omega_0 \vdash \ereturn \rightsquigarrow \ereturn\,v \dashv \Omega_n
    }

    \inferrule[E-IfThenElse-T]{
    \Omega \vdash \kop \rightsquigarrow \kw{true} \dashv \Omega'\\\\
    \Omega' \vdash s_1 \rightsquigarrow r \dashv \Omega''
    }{
      \Omega \vdash \eite{\kop}{s_1}{s_2} \rightsquigarrow r \dashv \Omega''
    }\fi

    \inferrule[E-Assign]{
      \Omega \vdash rv \rightsquigarrow v \dashv \Omega' \\
      \Omega'(p) \Rightarrow v_p \\\\
      v_p \text{ has no outer } \kw{loan}\text{s} \\
      x_\mathsf{old} \text{ fresh}\\\\
      \Omega'(p) \leftarrow v \Rightarrow \Omega'' \\
      \Omega''' = \Omega''[x_\mathsf{old} \mapsto v_p]
    }{
    \Omega \vdash \eassign p {rv} \rightsquigarrow () \dashv \Omega'''
    }\iflong

    \inferrule[E-Free]{
      \Omega(p) \Rightarrow \ebox v \\
      \Omega \vdash \eassign p \bot \rightsquigarrow () \dashv \Omega'
    }{
    \Omega \vdash \kw{free}\,p \rightsquigarrow () \dashv \Omega'
    }\fi

    \inferrule[E-Match]{
      \Omega \vdash p \stackrel s\Rightarrow C[\overrightarrow{f = v}] \\
      \Omega \vdash s \rightsquigarrow r \dashv \Omega'
    }{
    \Omega \vdash \ematch{p}{\ldots \mid C \to s \mid \ldots} \rightsquigarrow r \dashv \Omega'
    }\ifshort

    \inferrule[C-Shared-Borrow]{
      \ell' \text{ fresh} \\
      \esloan{\ell \cup \vec\ell}v \in \Omega \\\\
      \Omega' = \left[ \esloan{\ell \cup \ell' \cup \vec\ell}v
        \big/ \esloan{\ell \cup \vec\ell}v \right] \Omega
    }{
      \Omega \vdash \ecopy \esborrow\ell \Rightarrow
        \esborrow\ell' \dashv \Omega'
    }

    \inferrule[C-Shared-Loan]{
      \Omega \vdash \ecopy v \Rightarrow v' \dashv \Omega'
    }{
    \Omega \vdash \ecopy \esloan{\vec\ell} v \Rightarrow
        v' \dashv \Omega'
    }\fi
  \end{mathpar}
  \caption{Selected Reduction Rules for LLBC. We omit: \ifshort\textsc{E-IfThenElse-F}, \fi{}tuples (similar to
  constructor), sequences (trivial).  We also omit the handling of results -- these prevent further
  execution and simply get carried through. \iflong Boxes behave like regular ADT constructors, except for
  the \li+free+ Rust function, which receives primitive treatment, above.\fi}
  \label{fig:reduction}
\end{figure}

\begin{figure}
  \centering
  \smaller
  \begin{mathpar}
    \inferrule[R-Not-Shared]{
      \Omega(p) \Rightarrow v \\
      v \neq \esloan{\vec l}{v'}
    }{
      \Omega(p) \stackrel s\Rightarrow v
    }

    \inferrule[R-Shared]{
      \Omega(p) \Rightarrow \esloan{\vec l}{v}
    }{
      \Omega(p) \stackrel s\Rightarrow v
    }
  \end{mathpar}
  \caption{Auxiliary Judgment: Reading a Possibly Immutably-Shared Value. Rust
  allows matching on a value for which there are oustanding \emph{shared}
  borrows; the auxiliary $\stackrel s\Rightarrow$ read allows reading underneath
  a $\kw{loan}^s$\iflong, if applicable\fi.}
  \label{fig:match-read}
\end{figure}

\iflong
\begin{figure}
  \centering
  \smaller
  \begin{mathpar}
    \inferrule[C-Shared-Borrow]{
      \ell' \text{ fresh} \\
      \esloan{\ell \cup \vec\ell}v \in \Omega \\\\
      \Omega' = \left[ \esloan{\ell \cup \ell' \cup \vec\ell}v
        \big/ \esloan{\ell \cup \vec\ell}v \right] \Omega
    }{
      \Omega \vdash \ecopy \esborrow\ell \Rightarrow
        \esborrow\ell' \dashv \Omega'
    }

    \inferrule[C-Shared-Loan]{
      \Omega \vdash \ecopy v \Rightarrow v' \dashv \Omega'
    }{
    \Omega \vdash \ecopy \esloan{\vec\ell} v \Rightarrow
        v' \dashv \Omega'
    }

    \inferrule[C-Scalar]{
    v = \kw{true} \text{ or }\kw{false}\text{ or } n_\mathsf{i32} \text{ or } n_\mathsf{u32}\text{ or } \ldots
    }{
    \Omega \vdash \ecopy v \Rightarrow v \dashv \Omega
    }

    \inferrule[C-None]{
    }{
    \Omega \vdash \ecopy\enone \Rightarrow \enone \dashv \Omega
    }

    \inferrule[C-Some]{
      \Omega \vdash \ecopy v \Rightarrow v' \dashv \Omega'
    }{
    \Omega \vdash \ecopy {\esome v} \Rightarrow {\esome v'} \dashv \Omega'
    }

    \inferrule[C-Tuple]{
    \Omega_i \vdash \ecopy v_i \Rightarrow v'_i \dashv \Omega_{i+1}
    }{
    \Omega_0 \vdash \ecopy (\vec v) \Rightarrow (\vec v') \dashv \Omega_n
    }
  \end{mathpar}
  \caption{Auxiliary Judgment: Copying. We mimic MIR and see the copy of options
  and tuples as primitive operations. The judgment is undefined for any other
  construct as Rust's MIR only permits copying primitive data.}
  \label{fig:copy}
\end{figure}
\fi

\iflong
\begin{figure}
  \centering
  \smaller
  \begin{mathpar}
    \inferrule[A-Shorthand]{
      \ominus\; v_p
    }{
      v_p \text{ has no outer } \kw{loan}\text{s}
    }

    \inferrule[A-Scalar]{
      v = \kw{true} \text{ or }\kw{false}\text{ or } n_\mathsf{i32} \text{ or } n_\mathsf{u32}\text{ or } \ldots
    }{
      \ominus\; v
    }

    \inferrule[A-Tuple]{
      \ominus\; v_i
    }{
      \ominus\; (\vec v)
    }

    \inferrule[A-Constructor]{
      \ominus\; v_i
    }{
      \ominus\; C[\vec f = \vec v]
    }

    \inferrule[A-Borrow-M]{
    }{
    \ominus\; \emborrow\ell v
    }

    \inferrule[A-Borrow-R-S]{
    }{
    \ominus\; \kw{borrow}^{r,s} \ell
    }

    \inferrule[A-Bot]{
    }{
    \ominus\; \bot
    }
  \end{mathpar}
  \caption{Auxiliary Judgment: Absence of Outer Loans. We use shorthand notation $\ominus$ for this
  figure. Enforcing this
  criterion ensures that, at assignment-time, the memory we are about to write does not contained
  loaned-out data, as this would be unsound. This judgement is defined by omission, and is never
  valid for values of the form $\kw{loan}\;\_$. Such values, however, may appear underneath borrows,
  as the \textsc{A-Borrow-*} rules enforce no preconditions.}
  \label{fig:no-outer-loans}
\end{figure}
\fi

At the heart of our operational semantics is our treatment of borrows, which
captures ownership transfer.
We now introduce the core of our operational rules in \fref{reduction}, and describe in detail the
essential operations: borrows, moves, copies, and assignments.
Our rules start with \textsc{E-}, for evaluation rules.
Both the rules and our earlier syntax manipulate a
third flavor of borrows,
which we
dub ``\underline{r}eserved'' borrows, denoted as $\kw{borrow}^r$. A technical
device, they account for the two-phase borrows introduced by the Rust compiler in the
process of desugaring to MIR. We explain those later, in
\sref{semantics-reorganize}; they can be safely ignored for now.

\begin{figure}
  \centering
  \smaller
  \begin{mathpar}
    \inferrule[Write-G]{
    p = P[x] \\
    x \mapsto v_x \in \Omega \\\\
    \Omega \vdash p(v_x) \leftarrow v \stackrel g\Rightarrow v'_x \dashv \Omega' \\
    \Omega'' = \Omega'[x\mapsto v'_x]
    }{
    \Omega[p \mapsto v] = \Omega''
    }

    \inferrule[W-G-Shared-Borrow]{
    \esloan{\ell\cup \_}{v_p} \in \Omega \\
    \Omega \vdash p(v_p) \leftarrow v \stackrel g\Rightarrow v'_p \dashv \Omega' \\\\
    \Omega'' = \left[ \esloan{\ell\cup \_}{v'_p} \big/ \esloan{\ell\cup \_}{v_p} \right]\Omega'
    }{
    \Omega \vdash (\ederefs p)(\esborrow\ell) \leftarrow v \stackrel g\Rightarrow
    \esborrow\ell \dashv \Omega''
    }
  \end{mathpar}
  \caption{Auxiliary Judgment: Ghost Write. This judgment inherits all of the
  rules of the form \textsc{W-*}.}
  \label{fig:ghost-write}
\end{figure}

A few preliminary remarks about notation:
we rely on the auxiliary notion of a \emph{ghost update},
denoted $\Omega[p \mapsto v]$ -- this extends our earlier notation of $\Omega[x
\mapsto v]$ -- for updating an entry in the environment.
In contrast to run-time writes, previously introduced as $\Omega(p) \leftarrow v
\Rightarrow \Omega'$ (\Rule{Write}),
ghost updates do not have any effect at run-time (\fref{ghost-write}).
Instead, they allow us to perform the necessary book-keeping to statically keep
track of ownership and aliases.
The definition of ghost updates is almost identical to
run-time writes (\Rule{Write}), the only difference being that ghost updates can
follow a shared borrow to perform an administrative update underneath the shared
loan. As we will see shortly, this is leveraged by e.g.
\Rule{E-Shared-Or-Reserved-Borrow}, to track borrowing underneath a
shared borrow.
We write $\Omega\vdash \kop \rightsquigarrow v \dashv
\Omega'$ to indicate that in environment $\Omega$, operand $\kop$ reduces to $v$
and produces updated environment $\Omega'$. The $\rightsquigarrow$ judgment is
overloaded for other syntactic categories.
Finally, we write e.g. $\kw{loan}\not\in v$, with no arguments, to indicate that no kind
of loan should appear in $v$; we have similar syntactic conventions for other
restrictions.

For mutable borrows
(\Rule{E-Mut-Borrow}), we disallow: borrowing already-borrowed values (no
$\kw{loan}$); borrowing moved, uninitialized values (no $\bot$)
or reserved borrows; and borrowing
through a shared borrow (no $\ederefs\relax$ in p). This latter requirement refers to
the place $p$, not the value $v$ found at $p$: a value reachable via a shared
borrow is, inevitably, shared, and therefore cannot be uniquely owned by means of
a mutable borrow.
If these premises are
satisfied, we perform a ghost update of the environment, and mark $p$ as
loaned with identifier $\ell$.  The borrow evaluates to $\emborrow \ell v$,
which embodies unique ownership of value $v$ thanks to the exclusive loan $\ell$.

For immutable borrows (\Rule{E-Shared-Or-Reserved-Borrow}), we disallow moved or uninitialized
values (no $\bot$) and reserved borrows, but rule out mutable loans only: it is always legal in Rust to create
another shared borrow from a
value that has already been shared. The borrow evaluates to $\esborrow \ell$, a borrow without value
ownership. We need to record the fact that a fresh loan has been handed out; we
perform a ghost
update on the environment, to either augment the loan-set of the borrowed value
with $\ell$, or to introduce a new loan at $p$ to account for the fact that the
value at that place is now immutably borrowed.
The $r,s$ in the conclusion indicates that the rule may produce \emph{either} a
reserved or a shared borrow; again, it is safe to ignore the ``reserved''
variant for now.

For moves (\Rule{E-Move}), we disallow moving: $\bot$, already-borrowed values (no
loans), or reserved borrows. We also forbid moving \emph{through} a dereference. The former prevents
invalidating stray borrows; simply said, if a value has unterminated borrows, we
cannot obtain full ownership of it in order to perform the move. The latter
replicates Rust's constraint that no moves are allowed under a borrow.

For copies (\Rule{E-Copy}), we disallow copying mutable or reserved borrows, or
mutably-loaned values; we rely on an auxiliary judgment of the form $\Omega
\vdash \ecopy v \Rightarrow v' \dashv \Omega'$, meaning creating a copy of $v$ in
$\Omega$ produces $v'$ and returns a fresh environment $\Omega'$. This judgment
behaves %
like the earlier \Rule{Read}, except for shared borrows. When
copying a shared borrow, we automatically allocate a new loan
(\Rule{C-Shared-Borrow}) and augment the loan-set via a replacement (we use the
substitution notation) -- that is, we automatically perform a shared reborrow.
When copying a shared loan (\Rule{C-Shared-Loan}), we simply copy the actual
value without performing any shared-loan tracking; the ownership information
that regards the old value is irrelevant for the newly-copied value.

For matches (\Rule{E-Match}), we peek at the enum tag via
$\stackrel s\Rightarrow$; actual transfer of ownership with moves and
copies takes place in the suitable branch while executing $s$.
We note that in general, our \Rule{Read} judgment may return values of the form
$\kw{loan}^s$: this is useful e.g., to enforce that a value is not
loaned out, as in the premise of \Rule{E-Move}. For matches, however,
we merely need to read the enum tag; for this, we automatically dereference
shared loans via $\stackrel s\Rightarrow$.

All of our rules are in an explicit style: we prefer to add premises, rather
than rely on an implicit invariant by omission. For instance, we add many
premises to \Rule{E-Copy} rather than rely on the fact that the copy auxiliary
judgment has no rules for mutable borrows, reserved mutable borrows and
mutable loans. This also guides our lazy implementation of reorganizations.

We are now ready to define the semantics of assignments
(\Rule{E-Assign}). We reduce $rv$s first, and
remark that to obtain $v$, the various
rules for the $rv$ syntactic category must succeed. For instance, if the
right-hand side is a $\emove\relax$, then \Rule{E-Move} enforces all of its
preconditions. This means \Rule{E-Assign} operates with ownership of $v$, which
maps to our intuition for assignments in the presence of ownership and,
naturally, also corresponds to the Rust semantics.
What we do enforce, however, is that the value $v_p$ found at place $p$ should not
have any (outer) loans \iflong(\fref{no-outer-loans})\fi\ifshort\cite{longversion}\fi.
Overwriting a value that is currently loaned-out
would violate safety; we need to rule this out.
\label{no-outer-loans-discussion}
More precisely: loans may only appear behind pointer indirections; the value itself that
is being overwritten may not contain any loan.
The assignment rule ends by writing $v$ at the new place (using \Rule{Write}).
The rule for function call is identical, except it deals
with binding the arguments, locals and return variable.

One key point of \Rule{E-Assign} is that we retain the old value in the environment
$\Omega''$, under a fresh name $x_\mathsf{old}$ not accessible to the user-written program.
We call this retained value $x$ a \emph{ghost value}, because its only purpose is to avoid discarding useful ownership
knowledge; operationally, this is memory that can be actually reclaimed since it
isn't reachable anymore. Our third example at the beginning of the section
leverages this fact.

\subsection{Reorganizing Environments and Terminating Borrows}
\label{sec:semantics-reorganize}

We now present the final conceptual portion
of our operational semantics: reorganizing the
environment, which we used in our earlier examples to terminate borrows. We present rules in a
declarative style, to highlight the \emph{semantics} of Rust as opposed to the
\emph{implementation} of borrow-checking. A consequence of our declarative
approach is that we do not need to follow Rust's behavior to the letter; rather,
we reorganize borrows in a lazy fashion, and don't terminate a borrow unless we
need to get the borrowed value back. Concretely, our rules allow reorganization before
and after every statement (we have elided this from \fref{reduction} for
clarity). This has two concrete consequences. First,
we ignore the $\kw{drop}$ nodes from MIR -- indeed, they do not appear in
\fref{syntax}. Second, we claim that this captures a general semantics of
borrows; we substantiate that claim by showing, in \sref{evaluation}, how our
semantics can validate a Rust program checked with Polonius, an ongoing rewrite
of the borrow checker to allow for a larger class of Rust programs to be
accepted.

We define reorganizing via a set of rewriting rules that operate on the
environment $\Omega$ (\fref{reorg}). Since these rules are syntactic in nature, we rely on
value contexts $V[v]$, rather than our earlier semantic notions of
reads, writes and ghost updates. We omit administrative rules for
re-ordering environments at will. Our judgments are of the form $\Omega \hookrightarrow
\Omega'$, meaning $\Omega$ may be reorganized into $\Omega'$. We indulge in some
syntax overload; whenever used on the left-hand side of $\hookrightarrow$, we
understand $\Omega[x \mapsto v]$ to pattern-match on $\Omega$ to select a
mapping. This considerably simplifies notation.

\begin{figure}
  \centering
  \smaller
  \begin{mathpar}
    \inferrule[Not-Borrowed]{
      \not\exists V',V''.\ V[\cdot] = V'[\emborrow\_{(V''[\cdot])}] \\\\
      \not\exists V',V''.\ V[\cdot] = V'[\esloan\_{(V''[\cdot])}]
    }{
      \mathsf{not\_borrowed\_value}\,V
    }

    \inferrule[End-Shared-Or-Reserved-1]{
      \mathsf{not\_borrowed\_value}\,V
    }{
      \arraycolsep=1pt
      \begin{array}{lll}
        \Omega[x_1 \mapsto V[\kw{borrow}^{r,s} \ell], & x_2 \mapsto V'[\esloan{\ell}v]] &
          \hookrightarrow \cr
        \Omega[x_1 \mapsto V[\bot], & x_2 \mapsto V'[v]]
      \end{array}
    }

    \inferrule[Not-Shared]{
      \not\exists V',V''.\ V[\cdot] = V'[\esloan\_{(V''[\cdot])}]
    }{
      \mathsf{not\_shared\_value}\,V
    }

    \inferrule[End-Shared-Or-Reserved-2]{
      \mathsf{not\_borrowed\_value}\,V \\
      \ell \not\in \vec\ell
    }{
      \arraycolsep=1pt
      \begin{array}{lll}
        \Omega [x_1 \mapsto V[\kw{borrow}^{r,s} \ell], & x_2 \mapsto V'[\esloan{\ell\cup\vec\ell}v] ] &
          \hookrightarrow \cr
          \Omega[x_1 \mapsto V[\bot], & x_2 \mapsto V'[\esloan{\vec\ell}v]]
      \end{array}
    }

    \inferrule[End-Mut]{
      \{\kw{loan}, \kw{borrow}^r\} \not\in v \\
      \mathsf{not\_borrowed\_value}\,V
    }{
      \arraycolsep=1pt
      \begin{array}{lll}
        \Omega[x_1 \mapsto V[\emborrow\ell{v}], & x_2 \mapsto V'[\emloan\ell]]
          \hookrightarrow \cr
        \Omega[x_1 \mapsto V[\bot], & x_2 \mapsto V'[v]]
      \end{array}
    }

    \inferrule[Activate-Reserved]{
      \{\kw{loan}, \kw{borrow}^r\} \not\in v \\
      \mathsf{not\_shared\_value}\,V'
    }{
      \arraycolsep=1pt
      \begin{array}{lll}
        \Omega[x_1 \mapsto V[\eiborrow \ell], & x_2 \mapsto V'[\esloan{\ell}v]]
          \hookrightarrow \cr
        \Omega[x_1 \mapsto V[\emborrow\ell v], & x_2 \mapsto V'[\emloan\ell]]
      \end{array}
    }
  \end{mathpar}
  \caption{Reorganizing Environments}
  \label{fig:reorg}
\end{figure}

Our rules either render a value unusable ($\bot$), or strengthen it
($\kw{borrow}_m$, in the case of reserved borrows). For
these reasons, we demand unique ownership of the value in $V$ via
\Rule{Not-Borrowed}; that is, we can only terminate borrows for values that are
not themselves borrowed. Doing so, we precisely capture the constraints of
Rust with regards to reborrows. We now review the rules.

When ending a shared borrow, we render the borrow unusable henceforth, and
replace it with $\bot$. Then, two situations arise. If this is the last borrow
(\Rule{End-Shared-Or-Reserved-1}), i.e., if the loan-set is the singleton set $\left\{ \ell
\right\}$, we replace the shared loan with the previously-shared value. If there are more borrows out there
(\Rule{End-Shared-Or-Reserved-2}), we simply decrease the loan-set of the shared loan to
reflect that the borrow has been ended.

When ending a mutable borrow (\Rule{End-Mut}), we enforce that we own the value
we are about to return (i.e. not loaned).
The borrow then becomes unusable, and the borrowed value is returned to its rightful
owner.

This high-level approach to the Rust semantics allows us to very naturally
account for an oft-used Rust feature, namely two-phase borrows, which are introduced
in many places when desugaring to MIR. We account for those
through what we call \emph{reserved} borrows.
Reserved borrows are created just like shared borrows
(\Rule{E-Shared-Or-Reserved-Borrow}). However, reserved borrows cannot be
copied, dereferenced, or written into. Therefore, the
only way to use a reserved borrow is to strengthen it into a mutable borrow,
which is legal, as long as all other (shared or reserved) borrows have ended
(\Rule{Activate-Reserved}). Reserved borrows enable a variety of very common
idioms~\cite{rustc-dev} without resorting to more advanced desugarings.

These rules are declarative and non-ordered; in practice, our tool performs a
syntax-directed reorganization guided by the various preconditions on our rules.
For instance, whenever $\kw{loan}\not\in v$ appears as a premise, we perform a
traversal of $v$ to end whichever loans we encounter. This is another reason why
we prefer the ``explicit'' style of rules (i.e. with copious premises): they
clearly state expectations, and thus allow for a straightforward implementation.

\section{Symbolic Abstractions and Functional Translation}
\label{sec:translation}

Our semantics allows us to keep track of borrows and ownership in an exact
fashion.
We now ask: if we adopt a modular approach and treat function calls as
opaque, how much can we leverage borrows and regions to still enable precise
tracking of ownership and aliasing?
We answer that question with a
region-centric shape analysis that abstracts away the effect of a function
call on the ownership graph, via a notion of \emph{region abstraction}. We dub the
result our ``symbolic semantics''; it is, obviously, less precise than our
earlier concrete semantics; yet, it contains enough ownership and aliasing
information that we can generate a functional translation from it.
Our symbolic semantics very much resembles the concrete semantics; this
time, however, we turn out attention to the region information provided by
function signatures to
abstract away subsets of the ownership graph.

From this section onwards, we introduce a few additional restrictions on the
subset of Rust we can handle. Our concrete semantics supports loops, but our
symbolic semantics does not; we \emph{do}, however, support recursive functions. We disallow nested
borrows in function \emph{signatures}, but users can still manipulate
arbitrarily nested borrows within function bodies.
We also disallow instantiating a polymorphic function
with a type argument that contains a borrow.
Finally, we do not allow type declarations that contain borrows.
We believe most of these issues can be addressed with suitable amounts of
engineering; we discuss these limitations in detail in~\sref{future}.

\subsection{Symbolic Semantics by Example}
\label{sec:abstraction-examples}

\myparagraph{Symbolic Values; Matches}
A first concept we need to add to our toolkit is that of a symbolic
variable; that is, a variable whose type is known, but not its value: we write
$(\sigma: \tau)$. We now illustrate how symbolic variables
behave, notably in the presence of \li+match+es, which refine our static
knowledge about a symbolic variable.
From here on, we
make many constructions explicit so as to study a valid LLBC program;
importantly, $\kw{move}$s are now materialized.

\begin{minted}[mathescape]{rust}
fn f(mut o: Option<i32>) {  // $o \mapsto (\sigma: \kw{Option}\;\kw{i32})$
  let po = &mut o;          // $o \mapsto \emloan\ell; \quad po \mapsto \emborrow\ell{(\sigma: \kw{Option}\;\kw{i32})}$

  match *po {
    None => {               // $o \mapsto \emloan\ell; \quad po \mapsto \emborrow\ell{(\sigma: \kw{None})}$
      panic!() }

    Some => {               // $o \mapsto \emloan\ell; \quad po \mapsto \emborrow\ell{(\kw{Some}\;(\sigma': \kw{i32}))}$
      let r =
        &mut (*po).Some.0;  // $o \mapsto \emloan\ell; \quad po \mapsto \emborrow\ell{(\kw{Some}\;\emloan\ell')}; \quad o_\kw{ref} \mapsto \emborrow{\ell'}{(\sigma': \kw{i32})}$
      *r = 1; }};}          // $o \mapsto \emloan\ell; \quad po \mapsto \emborrow\ell{(\kw{Some}\;\emloan\ell')}; \quad o_\kw{ref} \mapsto \emborrow{\ell'}{1}$
\end{minted}

In the example above, $\sigma$ stands in for the function parameter whose
concrete value is unknown at run-time; $\sigma$ behaves like any other value
from our previous examples, and can be borrowed mutably (line 2).

A key requirement for the soundness of our semantics is to forbid changing
the enum variant of \li+o+, while its value or one of its fields is
borrowed: this disallows leftover borrows pointing to data of the wrong
(previous) type. We enforce this soundness criterion as follows. Assume, for the
sake of example, that the user at line 3 decides to mutate \li+o+, e.g. by doing
\li+o = None+. Our semantics for assignments looks up the symbolic value for the
left-hand side of the assignment (i.e., \li+o+), and demands that the symbolic
value have no oustanding ``outer'' loans (we formally define this criterion in
\iflong\fref{no-outer-loans}\fi\ifshort\cite{longversion}\fi). In order to satisfy this criterion, we must terminate
the borrow \li+po+ in order to obtain $o \mapsto \sigma; po \mapsto \bot$, which
then prevents any further use of \li+po+ -- we have successfully prevented a
type-incorrect usage.

At line 4, we perform a case analysis; at this stage, all we know is that the
scrutinee \li+*po+ evaluates to symbolic variable $\sigma$, of the correct type
$\kw{Option}\;\kw{i32}$. In order to check the branches, we treat each one of
them individually, in each case refining $\sigma$ with a more precise value
according to the constructor of the branch.
Simply said, in the \li+None+ case, we replace every occurrence of $\sigma$ with
$\kw{None}$, and in the \li+Some+ case, we replace every occurrence of $\sigma$
with $\kw{Some}\;(\sigma':\kw{i32})$, where $\sigma'$ is a fresh symbolic
variable.

More interesting pointer manipulations follow in the \li+Some+ branch. We borrow
the value within the option via \li+r+, using a projector syntax inspired by
MIR's internal representation of projectors.
This borrowing incurs no loss in precision in our
alias tracking: because we refined $\sigma$ earlier (in effect, performing a
strong-update of $\sigma$), we know that \emph{both} \li+o+ and \li+po+ are
unusable as long as \li+r+ lives.
More specifically, and in the vein of our remark above: should the user, for the
sake of example, decide to mutate via \li+po+, e.g. to change the enum variant
by doing \li+*po = None+ at line 10, our symbolic semantics would
give up ownership of \li+r+, in order to regain $po \mapsto
\emborrow\ell{(\esome1)}$, which by virtue of containing no ``outer'' loans
would make the update valid (see
\ref{no-outer-loans-discussion}, discussion of \Rule{E-Assign}).

\myparagraph{Function Calls: Single region case}
We now switch from the callee to the caller's perspective, and turn our
attention to function calls. We introduce a new concept of region abstraction to our borrow graph. An
abstraction owns borrows and loans, but does so abstractly; that is, we have
no aliasing information about values in an abstraction. Region abstractions
allow us to retain ownership and aliasing information in the presence of
function calls; they are introduced when a call takes place, upon which
they assume ownership of the call's arguments; they are
terminated whenever the caller relinquishes ownership of the return value, upon
which ownership flows back to the original arguments.

Before modifying the semantics from \sref{semantics}, we illustrate region abstractions
with an example. We revisit our earlier \li+test_choose+ function
(\sref{example:choose}).
\begin{minted}[mathescape,escapeinside=||]{rust}
let mut x = 0;   let mut y = 0;
let px = &mut x; let py = &mut y; // $ x \mapsto \emloan\ell_x,\  y \mapsto \emloan \ell_y,\  px \mapsto \emborrow{\ell_x}0,\  py \mapsto \emborrow{\ell_y}0$
let pz = choose(true, move px, move py);
// $x \mapsto \emloan\ell_x,\quad y \mapsto \emloan\ell_y,\quad px \mapsto \bot,\quad py \mapsto \bot,\quad pz \mapsto \emborrow{\ell_r}{(\sigma: \kw{uint32})},$ $\label{line:call:0}$
// $A(\rho) \ \{\quad \emborrow{\ell_x}0,\quad \emborrow{\ell_y}0,\quad \emloan\ell_r \quad \}$ $\label{line:call:1}$
*pz = *pz + 1; |\label{line:assign}|
// $x \mapsto \emloan\ell_x,\quad y \mapsto \emloan\ell_y,\quad px \mapsto \bot,\quad py \mapsto \bot,\quad pz \mapsto \emborrow{\ell_r}{(\sigma': \kw{uint32})},\;$         $\hspace{.9ex}$step 0$\label{line:step0:0}$
// $A(\rho) \ \{\quad \emborrow{\ell_x}0,\quad \emborrow{\ell_y}0,\quad \emloan\ell_r \quad \}$ $\label{line:step0:1}$
// $x \mapsto \emloan\ell_x,\quad y \mapsto \emloan\ell_y,\quad px \mapsto \bot,\quad py \mapsto \bot,\quad pz \mapsto \bot,$                             $\,$step 1$\label{line:step1:0}$
// $A(\rho) \ \{\quad \emborrow{\ell_x}0,\quad \emborrow{\ell_y}0,\quad \sigma' \quad \}$ $\label{line:step1:1}$
// $x \mapsto \emloan\ell_x,\quad y \mapsto \emloan\ell_y,\quad px \mapsto \bot,\quad py \mapsto \bot,\quad pz \mapsto \bot,$                             $\,$step 2$\label{line:step2:0}$
// $px' \mapsto \emborrow{\ell_x}{\sigma_x},\quad py' \mapsto \emborrow{\ell_y}{\sigma_y}$ $\label{line:step2:1}$
// $x \mapsto \sigma_x,\quad y \mapsto \emloan\ell_y,\quad px \mapsto \bot,\quad py \mapsto \bot,\quad pz \mapsto \bot,\quad px' \mapsto \bot,\quad py' \mapsto \emborrow{\ell_y}{\sigma_y}$   step 3$\label{line:step3}$
assert!(x == 1);
\end{minted}
Up to line 2, the usual set of rules apply and yield an environment that is
consistent with \sref{semantics}.
Our abstract rules come in at line 3, where we are
faced with a function call. We now need to abstract the call, that is, precisely
capture how the function call affects the borrow graph, without looking at the
definition of the function itself. To do so, we have only one piece of
information at our disposal: the type of \li+choose+, namely
{\small
$(\kw{bool}, \tmbrw\rho{\kw{uint32}}, \tmbrw\rho{\kw{uint32}}) \to
\tmbrw\rho{\kw{uint32}}$}.

The type of \li+choose+ conveys two key pieces of information: first, it
\emph{consumes} two mutable borrows in order to \emph{produce} a fresh (abstract)
return value; second, the borrows and the return value belong to the same %
\emph{region} $\rho$.
We proceed
as follows. We allocate a fresh region abstraction $A(\rho)$, which owns the consumed
arguments pertaining to region $\rho$; in our case, $\emborrow{\ell_x}0$ and
$\emborrow{\ell_y}0$.
 (In the case of
multiple regions per function type, we need to project the ownership of the
arguments along their respective regions; we handle this case formally in
§~\ref{sec:concrete-to-symbolic}.)
We know that the return value \li+pz+ has type $\tmbrw\rho{\kw{uint32}}$;
furthermore, the region in
the type tells us that the owner of this abstract value is the abstraction
$A(\rho)$. We perform a \emph{symbolic expansion} (also detailed in the next
section) to give \li+pz+ the shape
$\emborrow{\ell_r}{(\sigma: \kw{uint32})}$, pointing into a $\emloan\ell_r$ for
the return value owned by $A(\rho)$. We use $\sigma$ to denote a ``symbolic value''; such values
are not statically known, and receive a special treatment during the
translation.
We obtain the environment at lines ~\ref{line:call:0}-\ref{line:call:1}, where
the ownership of both \li+px+ and \li+py+ has been transferred to the
region abstraction; and where \li+pz+ has full ownership of a value loaned from the
region abstraction. Intuitively, a region abstraction is a bag containing borrows (what has
been consumed) and loans (what has been produced).

At line ~\ref{line:assign}, the mutation type-checks, and does not affect the abstract
environment: the symbolic value $\sigma$ borrowed through \li+pz+ is simply replaced
by a fresh symbolic value $\sigma'$ stemming from the addition.
At that stage, we cannot read from \li+x+ since it is mutably loaned; we
therefore need to reorganize the environment to make the assertion succeed.
Since we
do not have any precise knowledge about the aliasing relationship between
\li+x+, \li+y+ and \li+pz+, we cannot return ownership to \li+x+ directly; we
must return ownership \emph{en masse} by terminating region $A(\rho)$. We do so
by terminating the borrow for \li+pz+, which returns the abstract value
$\sigma'$ to
$A(\rho)$ (step 1, lines ~\ref{line:step1:0}-\ref{line:step1:1}).
Now that $A(\rho)$ has no outstanding loans left, we can terminate
$A(\rho)$ itself. This reintroduces in the environment borrows $l_x$ and $l_y$ with fresh
  values, and replaces the borrowed values they held ($0$ in both
  cases) with fresh
  symbolic values to account for potential modifications (lines ~\ref{line:step2:0}-\ref{line:step2:1}).
These borrows are ghost values, i.e. not
directly accessible to the user; they once again ensure we do not lose ownership
information.
A final reorganization of the environment
terminates $\ell_x$,
and makes \li+x+ usable again (line ~\ref{line:step3}).

\myparagraph{Multiple region case}
We now study a call to the \li+swap+ function, which permutes the two components
of a single tuple, located in two different regions.
\small
\[
\color{envcolor}
\kw{swap}:\ (z: (\tmbrw\alpha{\kw{uint32}}, \tmbrw\beta{\kw{uint32}})) \to
(\tmbrw\beta{\kw{uint32}}, \tmbrw\alpha{\kw{uint32}})
\]
\normalsize
We examine a call \li+let r = swap (move z)+ in the following environment:
\small
\[
\color{envcolor}
  x \mapsto \emloan\ell_x,\quad
  y \mapsto \emloan\ell_y,\quad
  z \mapsto (\emborrow{\ell_x}0, \emborrow{\ell_y}0)
\]
\normalsize
This time, the presence of two regions forces us to be more precise. We
introduce a new notion of projector, which comes in three flavors. We use an
\emph{input borrow projector} to dispatch each component of the argument to
its respective region abstraction. We use a \emph{loan projector} to dispatch each component
of the returned value to its respective region abstraction. And we use an
\emph{output borrow
projector} to determine the shape of the return value based on its type
information. (We write these $\eiproj\relax$, $\elproj\relax$ and
$\eoproj\relax$, respectively.)
Doing so, we rely on \emph{expansion rules} to
destruct and name the components of various tuples as needed. Thus, the
environment after the function call is:
\small
\[
\color{envcolor}
\begin{array}{l}
  x \mapsto \emloan\ell_x,\quad
  y \mapsto \emloan\ell_y,\quad
  z \mapsto \bot,\quad
  r \mapsto \ebproj(\sigma:
      (\tmbrw\beta{\kw{uint32}}, \tmbrw\alpha{\kw{uint32}})) \\
  A(\alpha) \{\\
    \quad \eiproj ((\emborrow{\ell_x}0, \emborrow{\ell_y}0):
      (\tmbrw\alpha{\kw{uint32}}, \tmbrw\beta{\kw{uint32}})) \\
    \quad \elproj (\sigma:
      (\tmbrw\beta{\kw{uint32}}, \tmbrw\alpha{\kw{uint32}})) \\
  \} \\
  A(\beta) \{\\
    \quad \eiproj ((\emborrow{\ell_x}0, \emborrow{\ell_y}0):
      (\tmbrw\alpha{\kw{uint32}}, \tmbrw\beta{\kw{uint32}})) \\
    \quad \elproj (\sigma:
      (\tmbrw\beta{\kw{uint32}}, \tmbrw\alpha{\kw{uint32}})) \\
  \}
\end{array}
\]
\normalsize
In the resulting environment, $z$ has been consumed. The return value $r$ needs
to be decomposed using an output borrow projector, according to its type. And
both region abstractions have the same content, which need to be projected along their
respective regions $\alpha$ and $\beta$.

A first, new reorganization rule (\fref{reorg-abstract})
allows us to refine the symbolic value $\sigma$ to be a tuple
$(\sigma_l, \sigma_r)$. Next, the two input projectors reduce based on the type: the left
component in $A(\alpha)$ remains, as it belongs to the enclosing region
$\alpha$, and the right component reduces to $\_$, an ignored value that does
not belong to $\alpha$. (The case of $\beta$ is symmetrical.)
The loan
projector behaves similarly and retains only the components pertaining to the
enclosing region; finally, the output borrow projector generates a pair of
borrows pointing to the corresponding abstract values. The resulting environment
is therefore as follows.
\small
\[
\color{envcolor}
\begin{array}{l}
  x \mapsto \emloan\ell_x,\quad y \mapsto \emloan\ell_y,\quad z \mapsto \bot,\quad
  r \mapsto (\emborrow{\ell_l}{\sigma_l},\ \emborrow{\ell_r}{\sigma_r}), \\[\medskipamount]
  A(\alpha) \{
    \quad (\emborrow{\ell_x}0, \_)
    \quad (\_,\ \emloan\ell_l)
  \quad \}, \\[\medskipamount]
  A(\beta) \{
  \quad (\_, \emborrow{\ell_y}0)
  \quad (\emloan\ell_r,\ \_)
  \quad \}\\[\medskipamount]
\end{array}
\]
\normalsize
\myparagraph{Discussion of examples}
We see our region abstractions as a form of magic wands; a function call consumes part
of the memory, and returns a magic wand (the region abstraction) along with its
argument (the returned value). Regaining ownership of the consumed memory
requires applying the magic wand to its argument, hence surrendering access to
the returned value.

Naturally, region abstractions set the stage for our functional translation:
introducing an abstraction translates to a call to a \emph{forward function},
while terminating an abstraction translates to a call to a \emph{backward
function}. But in order to get to the functional translation, we must first
define region abstractions more formally.

\subsection{From Concrete to Symbolic Semantics}
\label{sec:concrete-to-symbolic}

We define our symbolic semantics as an extension of our earlier formalism, along
with a new rule for function calls. First, we extend the value category to
account for symbolic values, denoted $\sigma$ (\fref{abstract-values}), as well as our three kinds
of projectors.

Following the intuition from our second example, above, we recall that the chief goal of
input and loan projectors is to distribute (``project''), at function call-time,
each component of a value to their corresponding abstraction region, while
output borrow projectors allow destructuring the caller's view of the return
value according to its type and region.
Projectors enjoy some duality: if we switch to the point of view of the callee,
output and loan projectors capture the fact that the effective arguments are ``on
loan'' from the callee's context.

We keep the syntactic overhead to a minimum. Our rules (and our implementation)
enforce strong invariants: for instance, input borrow and loan projectors may only
appear within region abstractions; and only a very restricted form of values may appear
under projectors.
But to keep
notation lightweight, we refrain from adding extra syntactic categories.
In the same spirit, we introduce some syntactic sugar. Whenever type
annotations are not needed, we skip $\tau$ in $(\sigma: \tau)$. Conversely,
whenever we need to access the type of a value (to make a region apparent), we
allow $(v: \tau)$ as a convenient way to bind the type. Finally, we start
leveraging region annotations in borrow types, and introduce a new form of
path $A(\rho)\mapsto$ to select an element from an abstraction.

\begin{figure}
  \centering
    \smaller
    \[
    \ifshort
    \arraycolsep=4pt
    \begin{array}{ll}
    \fi
    \begin{array}[t]{llll}
      v & ::=  \\
        && \dots & \text{(as before)} \\
        && (\sigma: \tau) & \text{symbolic value} \\
        && \eiproj v & \text{input borrow projector} \\
        && \elproj v & \text{loan projector} \\
        && \ebproj v & \text{output borrow projector} \\
        \ifshort
    \end{array}
    &
    \begin{array}[t]{llll}
    \fi\iflong
    \\[1ex]
  \fi

      \Omega & ::=  \\
        && \ldots & \text{(as before)} \\
        && A(\rho) \{ \vec v \}, \Omega \quad & \text{new region abstraction for \iflong region \fi}\rho \\
      \\[1ex]

      \rho & ::= & & \text{region identifier}
    \end{array}
    \ifshort
    \end{array}
    \fi
    \]
  \caption{Abstract Semantics: Environments, Values}
  \label{fig:abstract-values}
\end{figure}

\begin{figure}
  \centering
  \smaller
  \begin{mathpar}
    \inferrule[Decompose-Tuple]{
      \sigma_l, \sigma_r \text{ fresh}
    }{
      \Omega
      \overset\sigma{\underset{{(\sigma_l, \sigma_r)}}\hookrightarrow}
      \left[((\sigma_l, \sigma_r): (\tau_1, \tau_2))\Big/
      (\sigma: (\tau_1, \tau_2))\right]\Omega
    }

    \inferrule[Proj-Tuple]{
    }{
    \begin{array}{l}\Omega[A(\rho)\mapsto\kw{proj}_{\kw{in,l,out}}\,{(\sigma_l, \sigma_r)}] \hookrightarrow
    \cr\Omega[A(\rho)\mapsto(\kw{proj}_\kw{in,l,out}\,{\sigma_l}, \kw{proj}_\kw{in,l,out}\,{\sigma_r})]
    \end{array}
    }

    \inferrule[Proj-I-Mut-Match]{
    }{
    \begin{array}{l}\Omega[A(\rho)\mapsto\eiproj{(\emborrow\ell\sigma: \tmbrw\rho\tau)}] \hookrightarrow
    \cr\Omega[A(\rho)\mapsto\emborrow\ell(\sigma: \tau)]
    \end{array}
    }

    \inferrule[Proj-I-Mut-No-Match]{
    }{
    \begin{array}{l}\Omega[A(\rho)\mapsto\eiproj{(\emborrow\ell\_: \tmbrw\mu\tau)}] \hookrightarrow
    \cr\Omega[A(\rho)\mapsto\_]
    \end{array}
    }

    \inferrule[Proj-I-Shared-Match]{
    }{
    \begin{array}{l}\Omega[A(\rho)\mapsto\eiproj{(\esborrow\ell: \tbrw\rho\tau)}] \hookrightarrow
    \cr\Omega[A(\rho)\mapsto\esborrow\ell]
    \end{array}
    }

    \inferrule[Proj-I-Shared-No-Match]{
    }{
    \begin{array}{l}\Omega[A(\rho)\mapsto\eiproj{(\esborrow\ell: \tbrw\mu\tau)}] \hookrightarrow
    \cr\Omega[A(\rho)\mapsto\_]
    \end{array}
    }\iflong

    \inferrule[Proj-I-Symb]{
      \&\not\in\tau
    }{
    \Omega[A(\rho)\mapsto\eiproj{(\sigma: \tau)}] \hookrightarrow
    \Omega[A(\rho)\mapsto\sigma]
    }\fi

    \inferrule[Proj-Unfold-Mut-Match]{
      \sigma',\ell\text{ fresh}
    }{
    \arraycolsep=1.4pt
    \begin{array}{ll}\Omega[p\mapsto \eoproj(\sigma: \tmbrw\rho\tau),& A(\rho) \mapsto \elproj{\sigma}] \hookrightarrow
    \cr\Omega[p\mapsto \emborrow\ell\sigma',& A(\rho) \mapsto \emloan\ell]
    \end{array}
    }

    \inferrule[Proj-Unfold-Shared-Match]{
      \sigma',\ell\text{ fresh}
    }{
    \arraycolsep=1.4pt
    \begin{array}{ll}\Omega[p\mapsto \eoproj(\sigma: \tbrw\rho\tau),& A(\rho) \mapsto \elproj{\sigma}] \hookrightarrow
    \cr\Omega[p\mapsto \esborrow\ell,& A(\rho) \mapsto \esloan\ell\sigma']
    \end{array}
    }

    \inferrule[Proj-L-No-Match]{
      \&\not\in\tau
    }{
    \begin{array}{l}\Omega[A(\rho) \mapsto \elproj{(\sigma:\tau)}] \hookrightarrow
    \cr\Omega[A(\rho) \mapsto \_]
    \end{array}
    }

    \inferrule[End-Abstract-Mut]{
      \mathsf{not\_borrowed\_value}\,V
    }{
    \arraycolsep=1.4pt
    \begin{array}{ll}\Omega[ x \mapsto V[\emborrow\ell v], & A(\rho) \mapsto \emloan\ell ] \hookrightarrow
      \cr\Omega[ x \mapsto V[\bot], & A(\rho) \mapsto v ]
      \end{array}
    }

    \inferrule[End-Abstraction]{
      \kw{borrows}^m(A(\rho)) = \overrightarrow{\emborrow\ell\_} \\\\
      \kw{borrows}^s(A(\rho)) = \overrightarrow{\esborrow{\ell'}} \\\\
      \kw{loan},\elproj\not\in A(\rho).v_r \\
      \vec\sigma' \text{ fresh} \\
      \vec x_g, \vec y_g \text{ fresh}
    }{
    A(\rho),\Omega \hookrightarrow
    \Omega,
    \overrightarrow{x_g \mapsto \emborrow\ell\sigma'},
    \overrightarrow{y_g \mapsto \esborrow{\ell'}}
    }
  \end{mathpar}
  \caption{Reorganizing Environments with Abstract Values and Projectors}
  \label{fig:reorg-abstract}
\end{figure}

The new rewriting rules capture the behavior demonstrated with earlier examples.
A symbolic value may be decomposed structurally (\Rule{Decompose-Tuple}); we use
a substitution notation to indicate that there may be several occurrences of
$\sigma$, and all of them must be substituted at the same time. (Our second
example showcased this situation.) We tack the value $\sigma$ being destructed
and the pattern used for that purpose $(\sigma_l, \sigma_r)$ onto the arrow, so
that they can be conveniently recalled to generate a suitable let-binding in the
functional translation. Symmetrically, any kind of projector descends
along the structure of the symbolic value (\Rule{Proj-Tuple}).

  Input borrow projectors, once confronted with a borrow value, either discard it
  (as in either \Rule{Proj-I-Mut-No-Match} or \Rule{Proj-I-Shared-No-Match})
  or retain it
  (as in either \Rule{Proj-I-Mut-Match} or \Rule{Proj-I-Shared-Match}).
  As we forbid nested borrows for now, projections stop upon the first borrow they encounter;
  if the projector retains the borrow, we keep the whole borrowed value.

Loan and output borrow projectors reduce in lockstep: if the environment contains an
output borrow projector and a region abstraction contains a corresponding loan
projector over the same symbolic value $\sigma$, then we may turn them into a borrow and
a loan, respectively (\Rule{Proj-Unfold-Mut-Match}, \Rule{Proj-Unfold-Shared-Match}).
To give back ownership of a return value to the abstraction,
\Rule{End-Abstract-Mut} folds a mutable borrow back into an abstraction. Ending
a region abstraction itself is done via \Rule{End-Abstraction}; we use the
$\kw{borrow}$ notation to collect all the borrows, at once from all the values
in the abstraction. We require two things.
First, that the abstraction has no outstanding loans, i.e. the contents of
the abstraction are either symbolic values or borrows. Second, that the
abstraction contains no loan projectors. This latter precondition avoids
dangling output borrow projectors once the abstraction has disappeared.
Terminating the region abstraction returns ownership of the borrows, with fresh
symbolic values, to fresh (ghost) variables.

\subsection{From Symbolic Semantics to Functional Code}

At last, we explain how \aeneas, using the symbolic semantics,
generates a pure translation of the original LLBC program.
We make several hypotheses at this stage; beyond the restrictions we already
mentioned, we now assume every disjunction in
the control-flow is in terminal position. This is a strong restriction, which in
practice requires duplicating the continuation of conditionals and matches. This
is also a well-understood problem, known as ``computing a join'' in abstract
interpretation for shape analysis domains~\cite{rival2011abstract}, or the
``merge problem'' in Mezzo~\cite{protzenko2014mezzo}. Coupled with the fact that
problem space is highly constrained by Rust's lifetime discipline,
we are confident
that this can be addressed systematically and predictably.%

\iflong

\subsubsection{Translation Example: Forward Function}

As a starting example, we consider the forward translation of the
\li+call_choose+ function below, now presented in LLBC syntax with explicit
writes and moves, along with a fully-explicit return variable \li+x_ret+.
\begin{minted}[mathescape,escapeinside=||]{rust}
fn call_choose(mut p : (u32, u32)) -> (x_ret: u32) {
  let px = &mut p.0; |\label{line:synth1:px}|
  let py = &mut p.1; |\label{line:synth1:py}|
  let pz = choose(true, move px, move py); |\label{line:synth1:call}|
  *pz = *pz + 1; |\label{line:synth1:incr}|
  x_ret = move p.0; |\label{line:synth1:ret}|
  return;
}
\end{minted}
The translation of the forward function is carried out by performing a symbolic execution on
\li+call_choose+, and synthesizing an AST in parallel to reflect the effect of applying
the symbolic execution rules.

A key insight about our synthesis rules is that they are exclusively concerned
with \emph{symbolic values}; the actual variables from the source program
($x\mapsto\dots$) are mere book-keeping devices and have no relevance to the
translated program. Symbolic values $\sigma$, however, cannot be determined
statically; they thus compute at run-time, and as such are let-bound in the
target program.

We start the translation by initializing an environment, where \li+p+ maps to a symbolic
value $\sigma_0$. In parallel, we synthesize the function as an AST with a hole to be
progressively filled as we make progress throughout the translation.
We present both environment and translation side-by-side to reflect
the fact that they both make progress in parallel. We point, whenever
possible, to the specific rules that apply; these pointers can be skipped upon a
first reading, but might prove useful later, once the reader has encountered the
formal definition of the translation.

\medskip
\noindent
\begin{tabular}{p{.5\textwidth}p{.5\textwidth}}
\small
$
\color{envcolor}
\begin{array}{l}
  p \mapsto (\sigma_0 : (\kw{u32}, \kw{u32})) \quad
\end{array}
$
\normalsize
&
\begin{minipage}[t]{0.5\textwidth}
\begin{minted}{fstar}
let call_choose_fwd (s0 : (u32 & u32)) : u32 =
  [.]
\end{minted}
\end{minipage}
\end{tabular}

\medskip

\noindent
At line~\ref{line:synth1:px}, accessing field $0$ requires us to expand
$\sigma_0$; we first do so, which leads to the introduction of a let-binding in the
translation (\Rule{T-Destruct}, \Rule{Decompose-Tuple}).

\medskip
\noindent
\begin{tabular}{p{.5\textwidth}p{.5\textwidth}}
\small
$
\color{envcolor}
  p \mapsto (\sigma_1 : \kw{u32}, \sigma_2 : \kw{u32}) \quad
$
\normalsize
&
\begin{minipage}[t]{0.5\textwidth}
\begin{minted}{fstar}
let call_choose_fwd (s0 : (u32 & u32)) : u32 =
  let (s1, s2) = s0 in
  [.]
\end{minted}
\end{minipage}
\end{tabular}
\medskip

\noindent
The expansion above then allows us to evaluate the two mutable borrows on
lines~\ref{line:synth1:px}-\ref{line:synth1:py} via \Rule{E-Mut-Borrow}.
It is important to notice that borrows and assignments just lead to bookkeeping in the environment:
the synthesized translation is left unchanged.
We get the following:

\medskip
\noindent
\begin{tabular}{p{.5\textwidth}p{.5\textwidth}}
\small
$
\color{envcolor}
\begin{array}[t]{l}
  p \;\ \mapsto (\emloan\ell_1, \emloan\ell_2) \quad\\
  px \mapsto \emborrow{\ell_1}(\sigma_1 : \kw{u32}) \quad\\
  py \mapsto \emborrow{\ell_2}(\sigma_2 : \kw{u32}) \quad\\
\end{array}
$
\normalsize
&
\begin{minipage}[t]{0.5\textwidth}
\begin{minted}{fstar}
let call_choose_fwd (s0 : (u32 & u32)) : u32 =
  let (s1, s2) = s0 in
  [.]
\end{minted}
\end{minipage}
\end{tabular}
\medskip

\noindent
We then reach the function call at line~\ref{line:synth1:call}.
To account for the call, we do several things (\Rule{T-Call-Forward}).
As before (\sref{abstraction-examples}), we introduce a region abstraction to
account for \li+'a+; transfer ownership of the effective arguments to the
abstraction; then introduce a fresh symbolic value
$\sigma_3 : \tmbrw\alpha\kw{u32}$ to account for the returned value stored in \li+pz+.

Specifically, we transfer ownership of the effective arguments of the call to region $\alpha$, using
three input projections to retain only the borrows that pertain to $\alpha$.
The first argument gets projected as
$\eiproj\alpha\;\kw{bool}$, which reduces to $\_$ (irrelevant value). The next two arguments, each of type
$\tmbrw\alpha\kw{u32}$, reduce according to \Rule{Proj-I-Mut-Match}.

To handle the return value, we introduce a loan projector within the abstraction, and an output
borrow projector in the caller's context for \li+pz+; both refer to the same symbolic value
$\sigma_3$, meaning that they will both be refined simultaneously via the \textsc{Proj-Unfold-}
rules.

In parallel, we introduce a call to \li+choose_fwd+ in the synthesized translation.
As previously mentioned, the borrow types are translated to the identity; the input
arguments of \li+choose_fwd+ are thus simply \li+s1+ and \li+s2+.
Our translation is monadic, meaning we obtain:

\medskip
\noindent
\begin{tabular}{p{.5\textwidth}p{.5\textwidth}}
\small
$
\color{envcolor}
\begin{array}[t]{l}
 p \;\ \mapsto (\emloan\ell_1, \emloan\ell_2) \quad\\
 px \mapsto \bot \quad\\
 py \mapsto \bot \quad\\
 pz \mapsto \eoproj{(\sigma_3 : \tmbrw\alpha\kw{u32})} \quad\\
 A(\alpha) \{\\
 \quad \_,\\
 \quad \emborrow{\ell_1}(\sigma_1 : \kw{u32}),\\
 \quad \emborrow{\ell_2}(\sigma_2 : \kw{u32}),\\
 \quad \elproj{(\sigma_3 : \tmbrw\alpha\kw{u32}),}\\
 \}
\end{array}
$
\normalsize
&
\begin{minipage}[t]{0.5\textwidth}
\begin{minted}{fstar}
let call_choose_fwd (s0 : (u32 & u32)) : u32 =
  let (s1, s2) = s0 in
  s3 <-- choose_fwd true s1 s2;
  [.]
\end{minted}
\end{minipage}
\end{tabular}

\medskip
\noindent
Upon evaluating line~\ref{line:synth1:incr}, we expand $\sigma_3 : \tmbrw\alpha\kw{u32}$
to $\emborrow{\ell_3}(\sigma_4 : \kw{u32})$, following its type (\Rule{Proj-Unfold-Mut-Match}).
Once again, as the borrow types are translated to the identity, the expansion of $\sigma_3$
simply introduces a variable reassignment (\Rule{Pure-Mut-Borrow}).
This gives us:

\medskip
\noindent
\begin{tabular}{p{.5\textwidth}p{.5\textwidth}}
\small
$
\color{envcolor}
\begin{array}[t]{l}
  p \;\ \mapsto (\emloan\ell_1, \emloan\ell_2) \quad\\
  px \mapsto \bot \quad\\
  py \mapsto \bot \quad\\
  pz \mapsto \emborrow{\ell_3}(\sigma_4 : \kw{u32}) \quad\\
  A(\alpha) \{\\
  \quad \_,\\
  \quad \emborrow{\ell_1}(\sigma_1 : \kw{u32}),\\
  \quad \emborrow{\ell_2}(\sigma_2 : \kw{u32}),\\
  \quad \emloan\ell_3,\\
  \}
\end{array}
$
\normalsize
&
\begin{minipage}[t]{0.5\textwidth}
\begin{minted}{fstar}
let call_choose_fwd (s0 : (u32 & u32)) : u32 =
  let (s1, s2) = s0 in
  s3 <-- choose_fwd true s1 s2;
  let s4 = s3 in
  [.]
\end{minted}
\end{minipage}
\end{tabular}

\medskip
\noindent
Now that $\sigma_3$ has been decomposed, we can
symbolically execute the increment, which merely introduces a call to
\li+u32_add+ and generates a fresh variable $\sigma_5$ which stands for the unknown result of the
addition:

\medskip
\noindent
\begin{tabular}{p{.5\textwidth}p{.5\textwidth}}
\small
$
\color{envcolor}
\begin{array}[t]{l}
  p \;\ \mapsto (\emloan\ell_1, \emloan\ell_2) \quad\\
  px \mapsto \bot \quad\\
  py \mapsto \bot \quad\\
  pz \mapsto \emborrow{\ell_3}(\sigma_5 : \kw{u32}) \quad\\
  A(\alpha) \{\\
  \quad \_,\\
  \quad \emborrow{\ell_1}(\sigma_1 : \kw{u32}),\\
  \quad \emborrow{\ell_2}(\sigma_2 : \kw{u32}),\\
  \quad \emloan\ell_3,\\
  \}
\end{array}
$
\normalsize
&
\begin{minipage}[t]{0.5\textwidth}
\begin{minted}{fstar}
let call_choose_fwd (s0 : (u32 & u32)) : u32 =
  let (s1, s2) = s0 in
  s3 <-- choose_fwd true s1 s2;
  let s4 = s3 in
  s5 <-- u32_add s4 1;
  [.]
\end{minted}
\end{minipage}
\end{tabular}

\medskip
\noindent
Finally, the move at line~\ref{line:synth1:ret} requires retrieving the
ownership of \li+p.0+. Doing so requires ending the region abstraction
$A(\alpha)$ (introduced by the call
to \li+choose+), which in turns requires ending the loan inside the abstraction
(\Rule{End-Abstract-Mut}).
Followingly, we first end $\ell_3$ leading to the environment below,
where we notice $\sigma_5$ is the value
given back \emph{to} \li+choose+. Ending a loan (or a borrow, depending of
the point of view) leaves the synthesized code unchanged.

\medskip
\noindent
\begin{tabular}{p{.5\textwidth}p{.5\textwidth}}
\small
$
\color{envcolor}
\begin{array}[t]{l}
  p \;\ \mapsto (\emloan\ell_1, \emloan\ell_2) \quad\\
  px \mapsto \bot \quad\\
  py \mapsto \bot \quad\\
  pz \mapsto \bot \quad\\
  A(\alpha) \{\\
  \quad \_,\\
  \quad \emborrow{\ell_1}(\sigma_1 : \kw{u32}),\\
  \quad \emborrow{\ell_2}(\sigma_2 : \kw{u32}),\\
  \quad (\sigma_5 : \kw{u32})\\
  \}
\end{array}
$
\normalsize
&
\begin{minipage}[t]{0.5\textwidth}
\begin{minted}{fstar}
let call_choose_fwd (s0 : (u32 & u32)) : u32 =
  let (s1, s2) = s0 in
  s3 <-- choose_fwd true s1 s2;
  let s4 = s3 in
  s5 <-- u32_add s4 1;
  [.]
\end{minted}
\end{minipage}
\end{tabular}

\medskip
\noindent
Then, we actually end the region abstraction $A(\alpha)$ by moving back the
borrows $\ell_1$ and $\ell_2$ in the environment, with fresh symbolic
values $\sigma_6$ and $\sigma_7$ (\Rule{End-Abstraction}). Those are the values given back \emph{by} \li+choose+.
We mentioned earlier that ending a region abstraction translates as a call to
a backward function; we thus synthesize a call to \li+choose_back+ using
\Rule{T-Call-Backward}. The
\li+choose_back+ function
receives both the original arguments given to the call to \li+choose_fwd+ (that
is, \li+true+, \li+s1+ and \li+s2+, which
replicate the control-flow of the forward function) and the value
returned \emph{to} the region abstraction (that is, \li+s5+). The \li+choose_back+ function produces a
pair of values, namely those given back \emph{by} the region abstraction (that
is, \li+s6+ and \li+s7+).

\medskip
\noindent
\begin{tabular}{p{.5\textwidth}p{.5\textwidth}}
\small
$
\color{envcolor}
\begin{array}[t]{l}
  p \;\ \mapsto (\emloan\ell_1, \emloan\ell_2) \quad\\
  px \mapsto \bot \quad\\
  py \mapsto \bot \quad\\
  pz \mapsto \bot \quad\\
  px_g \mapsto \emborrow{\ell_1}(\sigma_6 : \kw{u32}) \quad\\
  py_g \mapsto \emborrow{\ell_2}(\sigma_7 : \kw{u32}) \quad\\
\end{array}
$
\normalsize
&
\begin{minipage}[t]{0.5\textwidth}
\begin{minted}{fstar}
let call_choose_fwd (s0 : (u32 & u32)) : u32 =
  let (s1, s2) = s0 in
  s3 <-- choose_fwd true s1 s2;
  let s4 = s3 in
  s5 <-- u32_add s4 1;
  (s6, s7) <-- choose_back true s1 s2 s5;
  [.]
\end{minted}
\end{minipage}
\end{tabular}

\medskip
\noindent
We can finally end borrow $\ell_1$ and evaluate the \li+return+, which ends the
translation (\Rule{T-Return-Forward}). As we save meta-information about the assignments to generate suitable
names for the variables, and also inline unnecessary let-bindings, \aeneas actually generates the
following function:
\begin{minted}{fstar}
let call_choose_fwd (p : (u32 & u32)) : u32 =
  let (px, py) = p in
  pz <-- choose_fwd true px py;
  pz0 <-- u32_add pz 1;
  (px0, _) <-- choose_back true px py pz0;
  Return px0
\end{minted}

\subsubsection{Translation Example: Backward Function \li+choose_back+}

We now proceed with the synthesis of the \emph{backward} translation of \li+choose+ from
section~\ref{sec:examples}. We recall its definition here for the sake of clarity.
\begin{minted}[mathescape,escapeinside=||]{rust}
fn choose<'a, T>(b : bool, x : &'a mut T, y : &'a mut T) -> &'a mut T {
  if b {|\label{line:synth2:if}|
    return x;|\label{line:synth2:returnx}|
  }
  else {
    return y;
  }
}
\end{minted}

\noindent
Unlike \li+call_choose+, the \li+choose+ function takes borrows as input parameters.
We thus need to track their provenance, from the point of view of the callee.
As a consequence, we initialize the environment by introducing an abstraction containing
loans so as to model the values owned
by the caller and \emph{loaned} to the function
for as long as $\alpha$ lives (\Rule{T-Fun-Backward}). This is the dual of the caller's point of view: the abstraction
contains loan projectors (that stand in for the effective arguments), and the context contains
output borrow projectors (that stand in for the formal parameters).

The synthesized function itself receives the same input parameters as the forward translation
of \li+choose+ (that is to say \li+b+, \li+x+ and \li+y+), together with an additional
parameter (\li+ret+) for the value which will be given back to the backward function upon ending
$\alpha$.

\medskip
\noindent
\begin{tabular}{p{.5\textwidth}p{.5\textwidth}}
\small
$
\color{envcolor}
\begin{array}[t]{l}
  A_\mathsf{input}(\alpha) \{
    \quad \_,
    \quad \emloan \ell_x,
    \quad \emloan \ell_y,
  \quad \}\\
  b \mapsto \sigma_b \quad\\
  x \mapsto \emborrow{\ell_x} (\sigma_x : \kw{t}) \quad\\
  y \mapsto \emborrow{\ell_y} (\sigma_y : \kw{t}) \quad\\
\end{array}
$
\normalsize
&
\begin{minipage}[t]{0.5\textwidth}
\begin{minted}{fstar}
let choose_back
  (t : Type) (b : bool) (x : t) (y : t)
  (ret : t) : result (t & t) =
  [.]
\end{minted}
\end{minipage}
\end{tabular}

\medskip
\noindent
We then evaluate the \li+if+ at line~\ref{line:synth2:if} (\Rule{T-IfThenElse}). This requires branching over symbolic
value $\sigma_b$.
We duplicate the environment and substitute $\sigma_b$ with $\ktrue$ for the first
branch, and $\kfalse$ for the second branch.
This substitution is a generic refinement which comes in handy for data types,
where it introduces binders for constructor arguments.
Of course, this introduces a branching in the synthesized translation.
Below, we show the environment for the first branch of the \li+if+:

\medskip
\noindent
\begin{tabular}{p{.5\textwidth}p{.5\textwidth}}
\small
$
\color{envcolor}
\begin{array}[t]{l}
  A_\mathsf{input}(\alpha) \{
    \quad \_,
    \quad \emloan \ell_x,
    \quad \emloan \ell_y,
  \quad \}\\
  b \mapsto \mathsf{true} \quad\\
  x \mapsto \emborrow{\ell_x} (\sigma_x : \kw{t}) \quad\\
  y \mapsto \emborrow{\ell_y} (\sigma_y : \kw{t}) \quad\\
\end{array}
$
\normalsize
&
\begin{minipage}[t]{0.5\textwidth}
\begin{minted}{fstar}
let choose_back
  (t : Type) (b : bool) (x : t) (y : t)
  (ret : t) : result (t & t) =
  if b then [.]
  else [.]
\end{minted}
\end{minipage}
\end{tabular}

\medskip
\noindent
We proceed with the evaluation of the first branch. Upon reaching
line~\ref{line:synth2:returnx}, we return \li+x+, that is to say the value
$\emborrow{\ell_x} (\sigma_x : \kw{t})$. This is the crucial part of the translation.
We first need to model the fact that after the return, the caller is free to use the
mutable borrow $\ell_x$ to perform in-place modifications until they end
region $\alpha$. Then, upon ending $\alpha$, we need to propagate the properly
updated values that were loaned to \li+choose+ at call site back to their original owners.
We proceed in two steps (\Rule{T-Return-Backward}).

We first replace $\sigma_x$ with $\sigma_\mathsf{ret}$, a symbolic value reserved for this purpose and
corresponding to the \li+ret+ parameter in the translation. We thus model the transfer of
ownership from the caller back to its callee, which happens when the caller ends
region $\alpha$. At this moment, the callee observes that $\sigma_x$ has been
updated to $\sigma_\mathsf{ret}$.
This gives us the environment:
\small
\[
\color{envcolor}
\begin{array}{l}
  A_\mathsf{input}(\alpha) \{
    \quad \_,
    \quad \emloan \ell_x,
    \quad \emloan \ell_y,
  \quad \}\\
  b \mapsto \kw{true} \quad\\
  x \mapsto \emborrow{\ell_x} (\sigma_\mathsf{ret} : \kw{t}) \quad\\
  y \mapsto \emborrow{\ell_y} (\sigma_y : \kw{t}) \quad\\
\end{array}
\]
\normalsize
For the second step, we end the loans in the input abstraction $A_\mathsf{input}$ to determine which
values are given back to the caller (in our case, the values borrowed by \li+x+ and \li+y+),
in place of the values loaned to \li+choose+ (and whose provenance is captured by
$\emloan \ell_x$ and $\emloan \ell_y$).
Ending those loans (\Rule{End-Abstract-Mut}) potentially leads to ending region
abstractions (\Rule{End-Abstraction}), introducing calls to
backward functions (this happens for \li+list_nth_mut+). We get the following environment:
\small
\[
\color{envcolor}
\begin{array}{l}
  A_\mathsf{input}(\alpha) \{
    \quad \_,
    \quad \sigma_\mathsf{ret},
    \quad \sigma_y,
  \quad \}\\
  b \mapsto \kw{true} \quad\\
  x \mapsto \bot \quad\\
  y \mapsto \bot \quad\\
\end{array}
\]
\normalsize

\noindent
From this environment we finally deduce the first branch of the translation:
\begin{minted}[mathescape,escapeinside=||]{fstar}
let choose_back (t : Type) (b : bool) (x : t) (y : t) (ret : t) : result (t & t) =
  if b then Return (ret, y)
  else [.]
\end{minted}

\noindent
We omit the translation of the second branch, which is similar.

\subsubsection{Synthesis Rules}
\fi %
The rules are in \fref{pleasenomorerules}, where \textsc{T-} stands for
translation;
they describe a process in
which we traverse the source program in a forward fashion, simultaneously
updating our symbolic environment and generating pure $\lambda$-terms.
Our final judgment
operates over LLBC statements $s$ and takes two forms. For statements that are
in terminal position, we write
$M, \Omega \vdash s \updownarrow e$, meaning statement $s$
compiles down to pure expression $e$ in environment $\Omega$ and translation
meta-data $M$. But for statements that are not in terminal position
(i.e., on the left-hand side of a semicolon), we are faced with the usual
mismatch between statement-based languages and let/expression-based languages.
We solve the issue by allowing expressions to contain a hole, to receive a
continuation -- we write $E[\cdot]$. Our judgement for non-terminal statements
is thus of the form $M, \Omega \vdash s \updownarrow E[\cdot] \dashv M', \Omega'$ -- we
note that this form produces an updated environment $M', \Omega'$ to allow chaining
with subsequent statements.
In practice, our
implementation uses continuation-passing style to keep things readable, as
opposed to an AST definition with holes for our target language.
In both cases, we let $\updownarrow$ be either $\downarrow$, for translating a
forward function, or $\uparrow^\rho$, for translating the backward function
associated to region $\rho$.
We spare the reader the grammar of expressions $e$, which
is a standard lambda-calculus; suffices to say that we use $\leftharpoonup$ to
denote the monadic bind operator, as in \sref{examples}; monadic returns appear
as $\kw{ret}$.
The synthesis of variables is
captured by the rules \textsc{Pure-*}; as we alluded to earlier, our translation
never encounters, nor produces, a source
variable $x$; rather, they structurally visit a symbolic value and map source
\emph{symbolic} variables to variables in the target $\lambda$-calculus
(\Rule{Pure-Symb}). Naturally, in practice, we use heuristics to pick sensible
names for the symbolic variables, thus guaranteeing that the output of our
translation is readable.
Conversion of types from source to target is almost the identity, except for
$\kw{Box}\,\tau$, $\tmbrw\rho\tau$ and $\tbrw\tau$ which become $\tau$,
consistently with \Rule{Pure-Box}, \Rule{Pure-Mut-Borrow} and
\Rule{Pure-Shared-Borrow}.

Another insight about our rules: in order to make progress, we may synthesize
fresh bindings at any time via a reorganization. For instance, if a symbolic
value with a tuple type is refined into the tuple of its components
(\Rule{Decompose-Tuple}), we need to mirror this fact in the generated program
(\Rule{T-Destruct}). We use a degenerate judgment of the form $M,\Omega\vdash
\emptyset \updownarrow\dots$, which appears in \Rule{T-Destruct},
\Rule{T-Reorg-Anytime} and \Rule{T-Call-Backward}.

The rest of the rules leverage our earlier concepts of region abstractions,
projections, and symbolic environments to precisely capture the relationship
between a function body and its parameters (callee), or a function application
and its arguments (caller). The rules for synthesis generally apply both for
generating forward and backward functions, with the exception of
\textsc{T-Return} and \textsc{T-Fun}.

We adopt the perspective of the caller and begin with function calls. In
\Rule{T-Call-Forward}, we follow the procedure outlined in our earlier examples.
First, we allocate a fresh $\sigma_r$ to stand in for the return value of the
call; we have one abstraction region per region in the function type. Each
abstraction region gains ownership of the relevant part of the function
arguments ($\eiproj{\vec v}$), and loans out whichever part of the symbolic
return value originates from that region ($\elproj\sigma_r$). These abstractions
augment the symbolic environment, along with an output projector for the return
value $\sigma_r$. We synthesize a monadic bind which introduces $\sigma_r$ in the generated
 program, and record in the meta-data $M$ that this
call happened with effective arguments $\vec v$.
Rule \Rule{T-Call-Backward} is not syntax-directed and may happen at any time;
in practice, we apply it lazily. We wish to terminate region $\rho$, associated
to an earlier function call found in $M$. We require a successful application of
\Rule{End-Abstraction}, so as to terminate abstraction region $A(\rho)$. This
returns ownership to us (the caller) of various borrows of symbolic values
$\vec\sigma'$. We thus need to synthesize a call to the backward function for
region $\rho$ of $f$, in order to compute in the translated code what is the
value of $\vec\sigma'$. The backward function receives the \emph{original}
effective arguments to $f$ (found in $M$), and the symbolic values that stands
for the terminated projector loans from $A(\rho)$.

We now switch to the perspective of the \emph{callee} and study function
definitions. We once again rely on region abstractions to explain the ownership
relationship between arguments and function body. In \Rule{T-Fun-Forward}, we
synthesize $f_\mathsf{fwd}$. In our initial environment, we have one abstraction
per region in the type of $f$; each abstraction region owns the parts of each argument
that are along region $\rho_i$; we execute the body in an environment where the
formal arguments $\vec x_\mathsf{arg}$ are each wrapped in an output projector.
The backward functions are synthesized in a similar way, except with extra
arguments.

Matches are delicate, and come in two flavors. We remark that our matches are
made up of non-nested, constructor patterns; by the time we examine Rust's
internal MIR language, nested patterns have already been desugared.

If our symbolic environment has
enough static knowledge to determine which particular branch of the data type we
are in (\Rule{T-Match-Concrete}), we do not bother with generating a trivial
match and simply generate code for the corresponding branch. If the scrutinee is
not known statically, then it is a symbolic variable
(\Rule{T-Match-Symbolic}).
For each branch $i$, we effectively perform a strong update of $\sigma$, replacing it with a
constructor value whose fields are themselves fresh symbolic variables (the
$\vec\sigma_i$). The same $\vec\sigma_i$ appear in the translated code (our target
lambda calculus \emph{can} bind arguments to constructors), thus preserving
proper lexical binding in the generated code. We remark that refining into a
constructor value is important for soundness -- lacking this precise
substructural tracking, we would lose precision in our borrow checking and
would allow type-incorrect programs.

We finally explain the rules for returning. In \Rule{T-Return-Forward}, the
value found in the special return variable dictates
what we return, so
long as we have ownership of this value. In \Rule{T-Return-Backward},
our goal is now to map sub-parts of $v_\mathsf{ret}$ back to their
original locations, using our region abstraction analysis. We ``symbolize''
$v_\mathsf{ret}$ using an auxiliary
\ifshort
function, elided
\fi\iflong
function (\fref{sym})
\fi; intuitively, $\kw{sym}$
replaces the sub-parts of $v_\mathsf{ret}$ (what we symbolically know about the
callee's return value $x_\mathsf{ret}$, so far) with matching sub-parts from $\sigma_\mathsf{ret}$
(the caller-updated return value, bound in \Rule{T-Fun-Backward}, provided at call site in
\Rule{T-Call-Backward}). For instance:
\smaller
\[\kw{sym}(\alpha, (\sigma_l, \sigma_r), (\emborrow\ell v: \tmbrw\alpha\tau,
\emborrow{\ell'} v': \tmbrw\beta\tau')) = (\emborrow\ell\sigma_l, \bot)\]
\normalsize
We then allow reorganizing the environment, so as to to allow destructuring
$\sigma_\mathsf{ret}$ suitably; this may also trigger calls to
\Rule{T-Call-Backward}, as is the case for \li+list_nth_back+. If the
region abstraction $A(\rho)$ is fully closed, the
values  it now holds determine the tuple we pass to the
monadic return.

\iflong
\newcommand\ksym{\kw{sym}}
\begin{figure}
  \centering
  \smaller
  \[
  \begin{array}[h]{lllr}
    \ksym(\alpha,(\vec\sigma),(\vec v)) & = & (\overrightarrow{\ksym(\alpha,\sigma,v)}) \\
    \ksym(\alpha,\sigma,\emborrow\ell v: \tmbrw\alpha\tau) & = & \emborrow\ell\sigma \\
    \ksym(\alpha,\_,\emborrow\ell v: \tmbrw{\beta}\tau) & = & \bot & \qquad \alpha \neq \beta
  \end{array}
  \]
  \caption{The $\kw{sym}$ function, used for the generation of backward
  functions. Specifically, $\kw{sym}(\alpha,\sigma,v)$ models the caller invoking
  the backward function for $\alpha$ with the value originally returned by the
  forward function to said caller. We abstract away the concrete view of the
  return value ($v$) into a symbolic view (modeled by $\sigma$) -- essentially
  saying that the caller may have arbitrarily mutated the return value while it
  owned it.}
  \label{fig:sym}
\end{figure}
\fi

Those are arguably the most important rules: we illustrate them with \li+choose+
(\sref{examples}).
In \li+choose_back+, $\sigma_\mathsf{ret}$ corresponds to the \li+ret+ argument.
In the \li+true+ branch, $v_\mathsf{ret}$ is $\emborrow{\ell_x}{\sigma_x}$.
Calling $\kw{sym}$ produces $\emborrow{\ell_x}{\sigma_\mathsf{ret}}$. We end
$\ell_x$, and thus propagate $\sigma_\mathsf{ret}$ back to the loan for
$\ell_x$. Besides, ending $\ell_y$ gives back the unchanged $\sigma_y$ and thus
closes the region abstraction $A(\alpha)$, which now contains $\left\{
\sigma_\mathsf{ret}, \sigma_y \right\}$; we return the symbolic values in the
exact same order, that is, \li+(ret, y)+.

\begin{figure}
  \centering
  \smaller
  \begin{mathpar}
    \inferrule[Pure-Mut-Borrow]{
      \Omega \vdash v \updownarrow e
    }{
      \Omega \vdash \emborrow\ell v \updownarrow e
    }

    \inferrule[Pure-Const]{
    }{
      \Omega \vdash n_\mathsf{i32} \updownarrow n_\mathsf{i32}
    }

    \inferrule[Pure-Tuple]{
      \Omega \vdash \vec v \updownarrow \vec e
    }{
      \Omega \vdash (\vec v) \updownarrow (\vec e)
    }

    \inferrule[Pure-Symb]{
    }{
      \Omega \vdash \sigma \updownarrow \sigma
    }

    \inferrule[Pure-Box]{
      \Omega \vdash v \updownarrow e
    }{
      \Omega \vdash \ebox v \updownarrow e
    }

    \inferrule[Pure-Shared-Borrow]{
    \esloan{\vec\ell}{v} \in \Omega \\ \ell \in \vec\ell \\\\
      \Omega \vdash v \updownarrow e
    }{
      \Omega \vdash \esborrow\ell \updownarrow e
    }

    \inferrule[T-Destruct]{
    \vec{\sigma'} \text{ fresh} \\
    M, \Omega \overset\sigma{\underset{(\vec{\sigma'})}\hookrightarrow} M', \Omega'
    }{
    M, \Omega \vdash \emptyset \updownarrow \kw{let}\;(\vec{\sigma'}) = \sigma\;\kw{in}\;[\cdot]\dashv M', \Omega''
    }

    \inferrule[T-Return-Forward]{
      \Omega_i \vdash \eassign{x_{\mathsf{local},i}}\bot \rightsquigarrow () \dashv \Omega_{i+1} \\
      \Omega_n (x_\mathsf{ret}) \Rightarrow v \\\\
      \Omega_n \vdash v \downarrow e\\
      \left\{ \kw{loan}, \bot, \kw{borrow}^r \right\} \not\in v
    }{
    M, \Omega_0 \vdash \kw{return} \downarrow e
    }

    \inferrule[T-IfThenElse]{
      M, \Omega \vdash \kop \updownarrow \sigma \dashv M', \Omega' \\
      \Omega' \overset\sigma{\underset{\kw{true}}\hookrightarrow}\Omega_1 \\
      \Omega' \overset\sigma{\underset{\kw{false}}\hookrightarrow}\Omega_2 \\\\
      M', \Omega_1 \vdash s_1 \updownarrow e_1 \\
      M', \Omega_2 \vdash s_2 \updownarrow e_2
    }{
      M, \Omega \vdash \eite{\kop}{s_1}{s_2} \updownarrow \eite{\sigma}{e_1}{e_2}
    }

    \inferrule[T-Seq]{
    M, \Omega \vdash s_1 \updownarrow E[\cdot] \dashv M', \Omega' \\\\
    M', \Omega' \vdash s_2 \updownarrow e \dashv M'', \Omega''
    }{
    M, \Omega \vdash s_1; s_2 \updownarrow E[e] \dashv M'', \Omega''
    }

    \inferrule[T-Call-Forward]{
    A(\rho_i) = \left\{ \overrightarrow{\eiproj v}, \elproj{\sigma_\kw{ret}} \right\} \\
    (\sigma : \tau) \in \vec v \text{ implies } \& \not\in \tau \\
    \vec\rho \text{ fresh} \\\\
    \sigma_\kw{ret} \text{ fresh} \\
    \Omega \vdash v_i \updownarrow e_i \\
    \Omega, \overrightarrow{A(\rho)} \vdash \eassign p {\eoproj\sigma_\kw{ret}} \rightsquigarrow () \dashv \Omega'
    }{
    M, \Omega \vdash \eassign{p}{f\langle\vec\rho\rangle\,\vec{v}} \updownarrow
    \sigma_\kw{ret} \leftharpoonup f_\mathsf{fwd}\,\vec e;\;[\cdot] \dashv (f\langle\vec\rho\rangle\,\vec e), M; \Omega'
    }

    \inferrule[T-Reorg-Anytime]{
    M, \Omega \vdash\emptyset\updownarrow E[\cdot] \dashv M', \Omega' \\\\
    M', \Omega' \vdash s \updownarrow E'[\cdot] \dashv M'', \Omega''
    }{
    M, \Omega \vdash s \updownarrow E[E'[\cdot]] \dashv M'', \Omega''
    }

    \inferrule[T-Call-Backward]{
      \Omega = A(\rho), \Omega' \\
      M = f\langle\ldots,\rho,\ldots\rangle\,\vec v, M' \\
      \Omega \vdash A(\rho).v_\kw{ret} \updownarrow e_\kw{ret} \\\\ %
      A(\rho), \Omega' \hookrightarrow
        \Omega'',
        \overrightarrow{x_g \mapsto \emborrow\ell\sigma'},
        \overrightarrow{y_g \mapsto \ldots} %
        \text{ \footnotesize via \Rule{End-Abstraction}}
    }{
    M, \Omega \vdash \emptyset \updownarrow
      (\vec \sigma') \leftharpoonup f_{\mathsf{back}(\rho)}\,\vec e \,e_\kw{ret}; [\cdot] \dashv M', \Omega''
    }

    \inferrule[T-Return-Backward]{
      \Omega_i \vdash \eassign{x_{\mathsf{local},i}}\bot \rightsquigarrow () \dashv \Omega_{i+1} \\\\
      \Omega_n (x_\mathsf{ret}) \Rightarrow v_\mathsf{ret} \\
      \left\{ \kw{loan}, \bot, \kw{borrow}^r \right\} \not\in v_\mathsf{ret} \\\\
      \Omega' = \Omega_n[x_\mathsf{ret} \mapsto \kw{sym}(\rho,
        \sigma_\mathsf{ret}, v_\mathsf{ret})] \\\\
      \Omega' \vdash \emptyset \uparrow^\rho E[\cdot] \dashv \Omega'' \\
      \Omega''[A(\rho)] = \left\{ \vec v \right\} \\\\
      \left\{ \kw{loan},\kw{proj} \right\}\not\in \vec v \\
      \Omega \vdash \vec v \uparrow \vec e
    }{
      \Omega_0 \vdash \kw{return} \uparrow^\rho E[\kw{ret}\;(\vec e)]
    }

    \inferrule[T-Fun-Backward]{
    A(\rho_i) = \left\{ \overrightarrow{\elproj{(\sigma: \tau})} \right\} \\
    \vec\sigma\text{ fresh}\\
    \emptyset,\Omega \vdash s \uparrow^\rho e \\\\
    \Omega = \overrightarrow{A(\rho)}, \overrightarrow{x\mapsto \eoproj\sigma},\,
    \vec x_\mathsf{local}\mapsto\vec\bot,\,
    x_\mathsf{ret}\mapsto\bot
    }{
    \begin{array}{l}
    \mathsf{fn}\,f\,\langle\vec\rho\rangle\,(\vec x: \vec\tau)\,
      (\vec x_\mathsf{local}: \vec \tau_\mathsf{local})\,(x_\mathsf{ret}: \tau_\mathsf{ret}) = s
    \uparrow^\rho
    \cr
      \kw{def}\;f_{\mathsf{back}(\rho)} =
      \lambda (\vec \sigma: \vec\tau)(\sigma_\mathsf{ret}: \etproj{\rho}{\tau_\mathsf{ret}}). e
    \end{array}
    }

    \inferrule[T-Fun-Forward]{
    A(\rho_i) = \left\{ \overrightarrow{\elproj{(\sigma: \tau})} \right\} \\
    \vec\sigma\text{ fresh}\\
    \emptyset,\Omega \vdash s \downarrow e \\\\
    \Omega = \overrightarrow{A(\rho)}, \overrightarrow{x\mapsto \eoproj\sigma},\,
    \vec x_\mathsf{local}\mapsto\vec\bot,\,
    x_\mathsf{ret}\mapsto\bot
    }{
    \begin{array}{l}
    \mathsf{fn}\,f\,\langle\vec\rho\rangle\,(\vec x: \vec\tau)\,
    (\vec x_\mathsf{local}: \vec \tau_\mathsf{local})\,(x_\mathsf{ret}: \tau_\mathsf{ret}) = s
      \downarrow \cr
      \kw{def}\;f_\mathsf{fwd} =
      \lambda (\vec \sigma: \vec\tau). e
    \end{array}
    }

    \inferrule[T-Match-Concrete]{
    \Omega(p) \stackrel s\Rightarrow C[\overrightarrow{f = v}] \\\\
    M,\Omega \vdash s \updownarrow e
    }{
    M, \Omega \vdash \ematch{p}{\ldots\mid C \to s\mid\ldots} \updownarrow
    e
    }

    \inferrule[T-Match-Symbolic]{
    \Omega(p) \stackrel s\Rightarrow (\sigma: t\,\vec\tau) \\
    \mathsf{type}\,t\;\vec\alpha = \overrightarrow{C [\vec f: \vec\tau_f]} \\
    \vec \sigma_i \text{ fresh}\\\\
    M, \left[ (C_i[\vec f_i: \vec \sigma_i]: t\,\vec\tau)\Big/ (\sigma: t\,\vec\tau) \right] \Omega \vdash s_i \updownarrow e_i
    }{
    M, \Omega \vdash \ematch{p}{\overrightarrow{C \to s}} \updownarrow
    \ematch\sigma{\overrightarrow{C\,\vec\sigma \to e}}
    }
  \end{mathpar}
  \caption{Functional Translation via our Symbolic Semantics}
  \label{fig:pleasenomorerules}
\end{figure}

\section{Implementing \aeneas}
\label{sec:implementation}

Our implementation is written in a mixture of Rust and OCaml. A first tool,
dubbed \charon, performs the translation from Rust's MIR internal representation
to LLBC. Concretely, \charon is a Rust compiler plugin that performs a large
amount of mundane, tedious tasks, such as: computing a dependency graph,
reordering definitions, grouping mutually recursive definitions together,
reconstructing data type creation, and generally getting rid of the idioms that
are definitely too low-level for LLBC (\sref{llbc}).
Once this is done, \charon dumps a JSON
file to disk containing the LLBC AST. We plan to switch to a more efficient binary
format in the future. \charon totals 9.5 kLoC (lines of code, excluding
whitespace and comments). \charon lives as a separate project because
we believe it has an existence of its own outside of \aeneas; we could easily
see other projects re-using a lot of the engineering work we performed in order
to share the implementation burden. We hope to present \charon to several other
tool authors in the near future.

\aeneas picks up the \charon-generated AST, and implements the transformations
described in \sref{semantics} and \sref{translation}. Practically speaking, we
have a single interpreter that runs in two modes, either concrete or symbolic.
The former produces a final value, if running a closed term; the latter produces
a translated program. We currently extract to \fstar, and have a Coq backend in
the works, along with plans for an HOL4 and a Lean backend.
Effectively, our symbolic interpreter acts as a borrow checker for Rust programs
using our semantic notion of borrows; we plan to investigate whether we can
isolate this checker to validate, e.g., bare C programs that would fit within
our admissible subset. We have written \aeneas in OCaml, a language much better
suited to the manipulation of ASTs than Rust. The implementation of \aeneas
totals 13.5kLoC.  Of those, 6kLoC are for the interpreters, and 4kLoC for the
translation and extraction. The rest contains library functions.

The implementation is, naturally, trusted. However, we have taken extraordinary care
to ensure that it is trustworthy. Notably, after every application of one of the
rules, we verify a large amount of invariants, such as: the environment is
well-typed; borrows are consistent; shared values don't contain a mutable loan,
etc. In practice, those invariants are extremely tight, and have led to the
great level of detail that our rules exhibit.
Should one turn off those invariant checks, the whole \charon + \aeneas invocation becomes almost
instantaneous, as opposed to a few seconds per file when constantly checking invariants.
In addition to those 13.5kLoC, we
have 6kLoC of comments; our implementation is truly written with great care.
Curious readers can find both \charon and \aeneas in the supplementary material.

\begin{figure}
\smaller
\begin{tabular}{llllllllll}
Project &
\mcrot{1}{l}{40}{General borrows} &
\mcrot{1}{l}{40}{Return borrows} &
\mcrot{1}{l}{40}{Loops} &
\mcrot{1}{l}{40}{Closures, traits} &
\mcrot{1}{l}{40}{Termination} &
\mcrot{1}{l}{40}{I/O} &
\mcrot{1}{l}{40}{Borrow check.} &
\mcrot{1}{l}{40}{Extrinsic} &
\mcrot{1}{l}{40}{Executable}\\
\aeneas & \checkmark & \checkmark & - & - & \checkmark & \checkmark & \checkmark & \checkmark & \checkmark \\
Electrolysis~\cite{electrolysis} & - & - & \checkmark & \checkmark & \checkmark & - & \checkmark & \checkmark & \checkmark \\
Creusot~\cite{creusot} & \checkmark & \checkmark & \checkmark & \checkmark & - & - & - & - & - \\
Prusti~\cite{prusti21} & \checkmark & \checkmark & \checkmark & \checkmark & - & - & - & - & - \\
\end{tabular}
\caption{\textbf{Comparison of Verification Frameworks Targeting Safe Rust}}
\label{fig:comparison}

\raggedright
\footnotesize General borrows: arbitrarily nested borrows and reborrows are allowed
in function bodies; Electrolysis supports a very restricted subset of such operations.
Return borrows: a function can return borrows; Electrolysis supports
a very restricted subset of such functions. Loop: the framework has support for loops.
Closures, traits: the framework has support for function pointers, closures and traits.
Termination: the framework allows to
reason about termination. I/O: we can reason about the external world like I/O.
Borrow check.: the framework doesn't trust nor need the Rust borrow checker; Prusti
extracts the lifetime information computed by the Rust borrow checker, then checks that those
lifetime are correct.
Extrinsic: the framework allows extrinsic proofs rather than intrinsic proofs; as stated, we believe
such proofs are
more amenable to modular reasoning and thus verification at scale.
Executable: the framework generates an executable translation of the Rust program.
\vspace{-2em}
\end{figure}

\myparagraph{Supported \aeneas Features}
\aeneas follows the design philosophy that no annotations should be added to the
Rust code; therefore, we have devised a few concrete mechanisms to make
integration of \aeneas-generated code easier.

Whenever translating a recursive function, \aeneas emits a \li+decreases+
annotation in the generated \fstar code. The annotation refers to a
yet-to-be-defined lemma that proves semantic
termination. This allows for a natural style, wherein the user invokes \aeneas
to generate e.g.  \li+Foo.fst+; the file references \li+Foo.Lemmas.fst+, which
is then filled out and maintained by the user. For Coq, we intend to rely on a
fuel parameter controlled by the user using a similar fashion.

We also provide support for interacting with the external world. Users of
\aeneas can choose to mark some modules in a given crate as ``opaque''; the
resulting functions appear as an interface file
only, meaning that they are \emph{assumed}. The user can then provide a
hand-maintained library of lemmas that characterize the behavior of these
external functions. We handle external dependencies in a similar fashion,
generating an interface which contains exactly those external functions and types
which are called from within the crate. This relieves the tool authors from having to maintain
wrappers for the entire Rust standard library.

Finally, we provide support for enhancing our working monad with an
external world. That is, rather than working in the error monad, we generate
code for a combined world and error monad. Combined with the opaque feature,
this allows the user to modularly state their assumptions about the world, and
gives us in practice a lightweight monadic effect system.

\section{Case studies}
\label{sec:evaluation}

\myparagraph{Hash Table}\label{section:hashtable}
To assess the efficiency of \aeneas as a verification platform, we study a
resizing hash table equipped with
\li+insert+, \li+get+ (immutable lookup), \li+get_mut+ (mutable lookup) and \li+remove+.
Each bucket is a linked list; \li+insert+ replaces the existing binding, if any;
resizing is automatic once a certain threshold is reached.
Our table is polymorphic in the type of the stored elements, but
right now, keys have a fixed type \li+usize+ (the proofs do not rely on
that). We plan to make the implementation generic over the key type once we
support traits.

\ifshort
We prove that our Rust hash table functionally behaves like a map. Most of the
challenges revolve around resizing after the threshold has been reached,
which involves reasoning about arithmetic overflow, mutable borrows for each
given slot, and element moves to avoid copies. We establish several invariants
about buckets and entries, then alternate between a semi-structural view (a
list of lists) and a high-level view (an associative list). Our insertion lemma
is as follows:
\fi

\iflong
The functional property we prove is that the hash table functionally behaves like a map.
In this regards, the interesting function is \li+insert+. This function delegates
insertion to \li+insert_no_resize+, then checks if the max load has been reached,
in which case it calls \li+try_resize+. We first prove that \li+insert_no_resize+
behaves as expected, that is to say: the binding for the inserted key maps to the new value,
and all the other bindings are preserved. We then prove that \li+try_resize+
is functionally the identity.
Interestingly, a function like \li+try_resize+ leverages many low-level
features of Rust: we have to be wary of arithmetic overflows when computing the new size,
we mutably borrow the slots vectors for in-place updates, and move elements so as
not to perform reallocations.

\begin{minted}{rust}
fn try_resize<'a>(&'a mut self) {
    // Check that we can resize (prevent overflows)
    let capacity = self.slots.len();
    let n = usize::MAX / 2;
    if capacity <= n / self.max_load_factor.0 {
        // Create a new table with a bigger capacity
        let mut ntable = HashTable::new_with_capacity(
            capacity * 2,
            self.max_load_factor.0,
            self.max_load_factor.1,
        );
        // Move the elements to the new table
        HashTable::move_elements(&mut ntable, &mut self.slots, 0);
        // Replace the current table with the new table
        self.slots = ntable.slots;
        self.max_load = ntable.max_load;
    }
}
\end{minted}

This proof is built on top of two invariants. First, in every slot,
all the keys are pairwise disjoint. This allows us to prove that \li+insert_no_resize+
is correct. Second, all the keys in a slot
have the proper hash. Combined with the first invariant, it gives us that
all the keys in the table are pairwise disjoint, which we need to prove that \li+try_resize+
is correct.
In combination with those invariants, stated in \li+hash_map_t_inv+,
we alternate between two high-level views of the hash
table. The first view is a list of lists, which corresponds to the low-level structure of the
hash table in rust. The second, which we use in the proof of \li+try_resize+, is a flattened
associative list.

Due to the low-level nature of Rust, precise arithmetic preconditions
pop up in the specification of the hash table. More specifically, inserting in the hash table doesn't
fail, if and only if the internal counter which tracks the number of entries doesn't overflow.
\fi

\begin{minted}{fstar}
val hash_map_insert_fwd_lem (#t : Type) (self : hash_map_t t) (key : usize) (value : t) :
  Lemma (requires (hash_map_t_inv self)) (ensures (
    match hash_map_insert_fwd t self key value with
    | Fail -> (* We fail only if: *)
      None? (find_s self key) /\ (* the key is not already in the map *)
      size_s self = usize_max (* and we can't increment `num_entries` *)
    | Return hm' -> (* In case of success: *)
      hash_map_t_inv hm' /\ (* The invariant is preserved *)
      find_s hm' key == Some value /\ (* [key] maps to [value] *)
      (forall k'. k' <> key ==> find_s hm' k' == find_s self k') /\ (* Other bindings unchanged *)
      (match find_s self key with (* The size is incremented, iff we inserted a new key *)
      | None -> size_s hm' = size_s self + 1
      | Some _ -> size_s hm' = size_s self)))
\end{minted}

By virtue of working with a (translated) pure program, we were able to focus on the
\emph{functional} behavior of the hash table and the important proof obligations,
such as the absence of arithmetic overflows, rather than memory reasoning.
Furthermore, the functional translation allowed us to prove the specifications
in an \emph{extrinsic} style, where we establish lemmas about the behavior of
existing functions, rather than in an
\emph{intrinsic} style, where we sprinkle the Rust code with assertions, calls
to auxiliary lemmas, and possibly materialize extra variables to aid
reasoning. We find that the extrinsic style not only yields a modular proof, but
also allows the control-flow of the proof to
refine the control-flow of the code. For instance, \li+insert_no_resize+
exhibits two logically different behaviors (key is present or not), even
though the code does not branch. Rather than materialize two
run-time code paths with different assertions, we perform the inversion
in the proof only.

The proofs took a total of 4 person-days, for an implementation of 201 LoC
without blanks and comments. We were hindered by some design choices of \fstar,
which generates proof obligations for the Z3 SMT solver; this mode of operation is
well-suited to intrinsic reasoning, but there is right now no way for the user
to have an interactive proof context like Coq. We are eager to investigate the
usability of \aeneas when coupled with backends that rely on
true interaction with tactics like Coq, and appear better suited for this
kind of extrinsic, functional proofs.

To the best of our knowledge, our implementation is the first verified hash
table in Rust. To obtain points of comparison, we therefore asked other non-Rust
verification experts or tool authors to estimate the effort to verify a hash
table using their respective frameworks. VST's hash table
exercise~\cite{appel2016verifiable} can be reasonably completed by students in
three days; this is, however, a non-resizing hash table. A hash table verified
using CFML~\cite{pottier2017verifying} required a week of work to establish
functional correctness; the original code, however, is in OCaml, which is a
higher-level language than C. Based on these comparisons, we conclude that
\aeneas is very competitive with other, more established verification
frameworks.

A final point of comparison is with other data structures in Low*, the subset of \fstar that compiles to C. A recent
paper~\cite{ho2021noise} verifies an imperative map using an associative linked
list. The authors reveal that this required several weeks of full-time work,
so we can confidently claim that we vastly improve the state of the art for
verifying low-level programs in \fstar.

\myparagraph{Non-Lexical Lifetimes (Polonius)}
We recently implemented a B$\epsilon$-tree~\cite{betree}, and ran it through
\aeneas; our verification efforts are ongoing.
In the process of doing so, we bumped into a limitation of the current Rust borrow checking
algorithm; interestingly, this limitation does not appear with \aeneas, owing to our semantic checking
of borrows.

\iflong
The \li+get_suffix_at_x+ function, below, looks for an element in a list, and returns a mutable borrow to
the suffix of this list starting at this element, for in-place modifications in a C-style
fashion. The current Rust borrow-checker, based on lexical lifetimes, is too
coarse to notice that the \li+Cons+ branch is valid. More specifically, it considers
that the reborrows performed through \li+hd+ and \li+tl+ should last until the end of the lexical
scope, that is, until the end
of the \li+Cons+ branch. We inserted the
error message printed by Rust in the comments.

Instead, one may notice that is it possible to end those borrows earlier,
after evaluating the conditional, in order to retrieve full
ownership of the value borrowed by \li+ls+ in the first branch of the \li+if+, and make the example
type-check. The ongoing replacement of the borrow-checker in Rust, named Polonius, implements a more
refined lifetime analysis, and accepts this program. However, the status of Polonius is unclear.

More interestingly, our semantic approach of borrows makes this program type-check without issues;
since our discipline is not lexical, but based on symbolic execution and a semantic approach to
loans, we accept the example without troubles, and are resilient to further syntactic tweaks of the program.
\fi
\ifshort
The function below exhibits what is known as a ``non-linear'' lifetime, in
which the borrow for \li+hd+ must be terminated in the \li+then+ branch in
order to regain full ownership of \li+ls+. The existing borrow checker of Rust
cannot account for this usage pattern, but an ongoing rewrite of the
borrow-checker, called Polonius, can. For \aeneas, there is no difference: our
precise semantics of borrows accounts of this use-case without trouble.
\fi

\begin{minted}{rust}
fn get_suffix_at_x<'a>(ls: &'a mut List<u32>, x: u32) -> &'a mut List<u32> {
    match ls {
        Nil => { ls }
        Cons(hd, tl) => { // first mutable borrow occurs here
          if *hd == x { ls // second mutable borrow occurs here
          } else { get_suffix_at_x(tl, x) } } } }
\end{minted}

\myparagraph{I/O and External Dependencies}
Real-world applications rely on external libraries and often need to interact
with the external world through I/O or sockets.
We elegantly model interaction with the outside environment using opaque modules, and
a state type that combines memory, IO and the outside world.
\iflong
In other words, \aeneas allows reasoning about such applications by lifting the generated code into a
combined-state + error monad, and relying on module signatures to model the interaction with
external functions.

When we designate a module as opaque, \aeneas treats all the
definitions coming from this module and reachable from the root module as opaque, and
generates an interface file accordingly, that is, with declarations
(\li+val+s), but no definitions (no \li+let+s).
The user is then free to provide models for those declarations, or simply state assumptions
by means of assumed lemmas.
This feature illustrates why a modular, type-directed translation like \aeneas'
is important: the user doesn't need to reveal
any information about the function's definition (or model its behavior using a
specification language); rather, the user can work post-translation in the
comfort of their favorite theorem prover.
Moreover, while it is possible to add annotations to function signatures in a local crate, this
possibility falls short when it comes to dealing with external dependencies,
over which the user has no control! Handling this oftentimes requires
tediously wrapping such dependencies in properly annotated modules. In contrast,
this work, and \charon and \aeneas, simply treat the external dependencies as opaque
by looking up the types and functions that are needed in the (non-opaque
modules of the) local crate, and generating corresponding declarations.

Finally, using a state-error monad allows us to introduce stateful reasoning when this is really
needed, for instance when the code uses I/O functions. The \li+state+ type, which models the external world,
is also an opaque type for which the user is free to provide a model or write assumptions, in a
fashion similar to opaque modules. In practice,
this gives us a lightweight effect system.

Let us illustrate those possibilities with the following example.
\fi
We set out to serialize our earlier hash table to the
disk. To account for this, we author \li+serialize+ and \li+deserialize+
functions in a separate opaque module outside of the scope of verification.
We
mark the module as opaque, meaning \aeneas generates the following
declarations.
\iflong
First, \li+insert_on_disk+, below, simply loads the map from the
disk, inserts a new entry, and stores the updated table back on disk.

\begin{minted}{rust}
fn insert_on_disk(key: Key, value: u64) {
    let mut hm = deserialize();
    hm.insert(key, value);
    serialize(hm); }
\end{minted}
\aeneas generates the following declarations to model the disk state and the
serialization and deserialization functions.
\fi
\begin{minted}{fstar}
type state : Type
val deserialize_fwd : state -> result (state & hash_map_t u64)
val serialize_fwd : hash_map_t u64 -> state -> result (state & unit)
\end{minted}
\iflong
Those definitions generate the following translation of \li+insert_on_disk+:
\begin{minted}{fstar}
let insert_on_disk_fwd (key : usize) (value : u64) : state -> result (state & unit) =
  hm <-- hashmap_utils_deserialize_fwd;
  hm <-- hash_map_insert_fwd_back u64 hm key value;
  _ <-- hashmap_utils_serialize_fwd hm;
  return ()
\end{minted}
\fi
\ifshort
We then hand-write salient lemmas about the two functions, e.g. serialization is
the inverse of parsing; because our monad now talks about the outside state, we can
precisely model the interaction of our parser/serializers with the outside
world.
\fi

\iflong
In order to reason about the function above, we define the (assumed) lemmas below,
which capture our assumption that once we store a table on the disk, all subsequent loads return
the same table, provided they succeed.

\begin{minted}{fstar}
val state_v : state -> hash_map_t u64

val serialize_lem (hm : hash_map_t u64) (st : state) : Lemma (
  match serialize_fwd hm st with | Fail -> True | Return (st', ()) -> state_v st' == hm)

val deserialize_lem (t : Type) (st : state) : Lemma (
  match deserialize_fwd st with
  | Fail -> True | Return (st', hm) -> hm == state_v st /\ state_v st' == state_v st)
\end{minted}

Relying on those lemmas, we can prove properties like the one below:

\begin{minted}{fstar}
val insert_on_disk_fwd (key : usize) (value : u64) (st : state) : Lemma (
  match hash_map_main_insert_on_disk_fwd key value st with
  | Fail -> True
  | Return (st', ()) ->
    let hm = state_v st' in
    match hash_map_insert_fwd_back u64 hm key value with
    | Fail -> False
    | Return hm' -> hm' == state_v st')
\end{minted}
\fi

\iflong
\myparagraph{Precise Reborrows (Discussion)}

Due to its precise management of borrows, we explained above that \aeneas can handle programs which
are supported by Polonius, and not by the current implementation of the borrow checker.
Because of the way we handle reborrows, there are actually cases of programs deemed invalid
even by Polonius, but supported by \aeneas.

For instance, in the example below, we first create a shared borrow that we store in
$pp$ at line~\ref{line:sreb:pp},
then a reborrow of a subvalue borrowed by $pp$ that we store in $px$ at line~\ref{line:sreb:px}.
Upon evaluating the assignment at line~\ref{line:sreb:assign}, $px$ simply maps
to a shared borrow of the first component of $p$ (environment at
lines~\ref{line:sreb:env0}-\ref{line:sreb:env1}).
Importantly, even though $px$ reborrows part of $pp$, there are no links
between $px$ and $pp$: our semantics does not
track the hierarchy between borrows and their subsequent reborrows.
In other words, line~\ref{line:sreb:px} is equivalent to \li+let px = &p.0+, where we borrow
directly from $p$ without resorting to $pp$.
This implies that, upon ending the borrow $\ell_p$ stored in $pp$ at
line~\ref{line:sreb:assign}, we do not need to end $\ell_x$ stored in $px$,
which in turn allows us to legally evaluate the assertion at line~\ref{line:sreb:assert}.

\begin{minted}[mathescape,escapeinside=||]{rust}
let mut p = (0, 1);
let pp = &p; |\label{line:sreb:pp}|
// $ p \cspace \mapsto \esloan{\ell_p}{(0,\ 1)} $
// $ pp \mapsto \esborrow\ell_p $
let px = &(*pp.0); |\label{line:sreb:px}|
// $ p  \cspace \mapsto \esloan{\ell_p}{(\esloan{\ell_x}{0},\ 1)} $ $\label{line:sreb:env0}$
// $ pp \mapsto \esborrow\ell_p $
// $ px \mapsto \esborrow\ell_x $ $\label{line:sreb:env1}$
p.1 = 2; |\label{line:sreb:assign}|
// $ p  \cspace \mapsto (\esloan{\ell_x}{0},\ 2) $
// $ pp \mapsto \bot $
// $ px \mapsto \esborrow\ell_x $
assert!(*px == 1); |\label{line:sreb:assert}|
\end{minted}

\noindent
When we attempt to borrow check this program (with Polonius or the current implementation
of the borrow checker), the Rust borrow checker considers that $px$ reborrows $pp$, and
thus needs to end before $pp$ ends. It consequently fails with the following error message:

\begin{minted}[linenos=false]{rust}
cannot assign to p.1 because it is borrowed
   |
 2 |     let pp = &p;
   |              -- borrow of p.1 occurs here
 5 |     let px = &(pp.0);
 9 |     p.1 = 2;
   |     ^^^^^^^ assignment to borrowed p.1 occurs here
13 |     assert!(*px == 1);
   |             --- borrow later used here
\end{minted}

The code snippet below illustrates a similar example with mutable borrows.
We create a borrow $px1$ of (the value of) $x$ at line~\ref{line:mreb:px1}, then reborrow
this value through $px2$ at line~\ref{line:mreb:px2}. At line~\ref{line:mreb:update},
we then update $px1$ to borrow $y$. At this point, $px2$ still borrows $x$.
The important point to notice is that upon performing this update, we remember
the old value of $px1$ in the ghost variable $px1_{\mathsf{old}}$ to not lose information
about the borrow graph (environment at lines~\ref{line:mreb:env0}-\ref{line:mreb:env1}).
Similarly to the previous example with shared borrows,
the resulting environment doesn't track
the fact that $px2$ was created by reborrowing the value initially borrowed by $px1$:
there are no links between those two variables.
Consequently, upon ending borrow $\ell_y$ (stored in $px1$) to access
$y$ at line~\ref{line:mreb:asserty}, we don't need to end $\ell_2$ (stored in $px2$).
This in return allows us to legally derefence $px2$ at line~\ref{line:mreb:assertpy2}.

\begin{minted}[mathescape,escapeinside=||]{rust}
let mut x = 0;
let mut px1 = &mut x; |\label{line:mreb:px1}|
let px2 = &mut (*px1); // Reborrow: px2 now borrows (the value of) x $\label{line:mreb:px2}$
// $ x \cspace\enspace \mapsto \emloan\ell_1 $
// $ px1 \mapsto \emborrow{\ell_1}{(\emloan\ell_2)} $
// $ px2 \mapsto \emborrow{\ell_2}{0} $
let mut y = 1;
px1 = &mut y; // Update px1 to borrow y instead of x $\label{line:mreb:update}$
// $ x \cspace\enspace\;\;\;\, \mapsto \emloan\ell_1 $ $\label{line:mreb:env0}$
// $ px1_{\mathsf{old}} \mapsto \emborrow{\ell_1}{(\emloan\ell_2)} $
// $ px2 \;\;\;\; \mapsto \emborrow{\ell_2}{0} $
// $ y \cspace\enspace\;\;\;\, \mapsto \emloan\ell_y $
// $ px1 \;\;\;\; \mapsto \emborrow{\ell_y}{1} $ $\label{line:mreb:env1}$
assert!(*px1 == 1);
assert!(*px2 == 0);
assert!(y == 1); // End the borrow of y through px1 (shouldn't impact px2!) $\label{line:mreb:asserty}$
// $ x \cspace\enspace\;\;\;\, \mapsto \emloan\ell_1 $
// $ px1_{\mathsf{old}} \mapsto \emborrow{\ell_1}{(\emloan\ell_2)} $
// $ px2 \;\;\;\; \mapsto \emborrow{\ell_2}{0} $
// $ y \cspace\enspace\;\;\;\, \mapsto 1 $
// $ px1 \;\;\;\; \mapsto \bot $
assert!(*px2 == 0); // Considered invalid by rustc, but accepted by Aeneas $\label{line:mreb:assertpy2}$
\end{minted}

\noindent
When attempting to borrow check this code snippet, we get the following error:

\begin{minted}[linenos=false]{rust}
cannot use y because it was mutably borrowed
   |
 8 |     px1 = &mut y; // Update px1 to borrow y instead of x
   |           ------ borrow of y occurs here
...
16 |     assert!(y == 1); // End the borrow of y through px1 (shouldn't impact px2!)
   |             ^ use of borrowed y
22 |     assert!(*px2 == 0); // Considered invalid by rustc, but accepted by Aeneas
   |             ---- borrow later used here
\end{minted}

The two examples above exemplify cases where the Rust borrow checker deems a program
as invalid, while \aeneas accepts it.
We do not claim that this is a strong limitation of the Rust borrow checker: these use cases
seem quite anecdotal and are probably useless in practice.
However, we believe the ability of \aeneas to precisely capture the behavior of such
use cases supports our claim that our semantics really captures the essence
of the borrow mechanism.

\fi

\ifshort
\section{Future work; related work; conclusion}
\label{sec:future}
\label{sec:conclusion}
\myparagraph{Future work}
\fi
\iflong
\section{Future work}
\label{sec:future}
\fi
We have admitted many of \aeneas' current limitations through this paper. We
plan to address loops and disjunctions in the control-flow as soon as possible,
using the techniques we referenced earlier. Doing so should bring
us closer to feature-parity with Creusot and Prusti. Next are traits,
our Coq backend, and large-scale use-cases. We believe all of the above to be
engineering tasks; the semantic insights are in this paper.

Our formalization provides a precise semantics of ownership in Rust; as we
alluded to earlier, we can explain not only extensions of the Rust
borrow-checker (Polonius), but trickier programs that Polonius
cannot yet account for.
Our next unit of work is a proof of soundness of our semantics and translation, possibly
against Stacked Borrows and/or RustBelt. Doing so would establish \aeneas as
an alternate borrow checker for Rust, possibly informing future evolutions.
\iflong

We now review in detail the remaining restrictions about the subset of Rust we
can handle.
We believe function pointers and closures fit naturally within our approach. Our reasoning is
entirely type-based, so a function can be fully characterized by a forward function
and a list of backward functions, one for each region variable (lifetime).
From there, adding support for traits should mostly be a matter of engineering.

Our current restriction about nested borrows in function types is overly conservative, and aims
to simplify and unambiguously rule out the truly difficult case, which is that of
nested mutable borrows living in different regions, e.g., \li+&'a mut &'b mut T+.
Manipulating such functions is made difficult by
operations like what we dub \emph{borrow overwrites}.
Borrow overwrites happen when we update an \emph{inner} borrow so as to borrow
a different value, as exemplified by line~\ref{line:nested:overwrite} in the example below.
Such operations are hard to account for in our abstract, symbolic semantics.

We extended our rules to handle such cases, but testing those in our
implementation revealed they were slightly imprecise.
More specifically and as a teaser, \aeneas was able to perform a symbolic execution
on the example below \emph{without} line~\ref{line:nested:chained}. However,
adding the chained function call would eventually lead to a state where the interpreter is
unable to legally apply any rule because the borrow graph is too imprecise,
making it get stuck and fail.
Further investigation seemed to indicate that our region abstractions are
too monolithic and need to be split into \emph{subregion} abstractions.
We will investigate this solution in future work.

\begin{minted}[mathescape,escapeinside=||]{rust}
fn f<'a, 'b>(ppx : &'a mut &'b mut u32) : &'a mut &'b mut u32;

let mut x = 0;
let mut px = &mut x;
let ppx = &mut px;
let mppx = &mut (*ppx);
let ppy = f(move mppx); // First function call
**ppy = 1;
let mut y = 2;
*ppy = &mut y; // Borrow overwrite - hard to track because ppy was returned by a function call $\label{line:nested:overwrite}$
let mppy = &mut (*ppy);
let ppz = f(&move mppy); // Chained function call $\label{line:nested:chained}$
\end{minted}

For the time being, having support for nested mutable borrows doesn't seem to be
extremely relevant in practice, with regards to the expressivity of the subset we support.
It often happens that Rust programmers manipulate
functions which receive as input a mutable borrow of a structure which contains itself mutable borrows.
However, such functions generally do not perform borrow overwrites, in which case it is possible to
split such nested borrows into non-nested borrows, i.e., split the input parameter into several input parameters.
In other words, the semantic expressivity provided by nested mutable borrows doesn't seem to be exploited
much by Rust developers. We plan to investigate if there are actual use cases which rely on this feature.

Our lack of support for borrows within ADTs mostly comes from engineering concerns. When an ADT contains
borrows, ending those borrows requires to give back \emph{exactly} those values which were
borrowed. This sometimes requires introducing some ``backward'' type definitions, which only
contain such fields. We forbid the instantiation of type parameters with types containing borrows
for the same reasons, and to control the introduction of nested borrows in a conservative manner.
Moreoever, instantiating a type parameter with a type containing mutable borrows transforms
any function into a function which requires the generation of backward translations. This can be
addressed by introducing a ``generic'' backward function for every
type parameter. In short, this would require generating higher-order backward functions.
Again, doing so will require a fair amount of engineering time.

\fi%
A final unit of future work is to strengthen the \charon tool; many authors of
Rust verification tools seem to re-implement comparable compiler plugins.
Sharing engineering efforts can only benefit the wider community.

\ifshort
\myparagraph{Related work}
\fi
\iflong
\section{Related work; conclusion}
\label{sec:conclusion}
\fi
Electrolysis~\cite{electrolysis} most resembles \aeneas. The tool translates
Rust programs to a pure lambda-calculus, then targets the
Lean~\cite{moura2015lean} proof assistant. It relies on lenses to model mutable
borrows, and as such comes with severe restrictions (\fref{comparison}); for
instance, functions may only return borrows to their first argument.
Electrolysis does not come with a formal model, and thus does not make
a case for semantic correctness. As such, it resembles a very pragmatic
``transpiler'' rather than a compiler; for instance, traits map to type classes,
because they, at a high-level, work in a similar fashion.

RustBelt~\cite{jung2017rustbelt} targets a different problem than \aeneas:
proving the soundness of Rust's type system, and proving the correctness of
unsafe Rust programs using the Iris framework~\cite{jung2018iris}. RustBelt is
an impressive framework and allows composing safe code with unsafe code using a
notion of semantic typing. We see RustBelt as the exact complement of \aeneas:
RustBelt allows proving fiendishly difficult, small pieces of unsafe Rust code,
while \aeneas allows reasoning about large amounts of safe Rust, without
resorting to a full-fledged framework like Iris.

RustHorn~\cite{matsushita2020rusthorn} uses a device called
\emph{prophecy variables} to generate a pure, non-executable logical encoding of
Rust programs. The original paper contains a non-mechanized proof that their
logical encoding is sound with respects to a memory-based semantics of the
original Rust program. RustHornBelt~\cite{RustHornBelt} uses the RustBelt
framework to mechanically prove the soundness of the RustHorn-style logical
encoding. We plan to investigate using this style of proof to mechanically
establish the soundness of our ownership-centric semantics.

Creusot~\cite{creusot} is a tool that builds on RustHornBelt to generate proof
obligations that can then be discharged to SMT; the authors
have a proof of soundness formalized in Coq for a simplified model of MIR.
Their design chooses \emph{automated},
\emph{intrinsic} proofs: they introduce an annotation language for specifications,
wrap a large part of the standard library in it, then rely on requires/ensures
clauses and annotations to perform the proofs. This style emphasizes a logical
encoding as opposed to an executable specification, and a closed-world approach
where the verified code cannot naturally be integrated as part of, e.g., a large
Coq development. The advantage of this style is that they can easily require
annotations for, e.g., loop invariants.

Prusti's frontend~\cite{prusti} is very similar to Creusot's.
The tool uses Rust's type system to guide the application of rules
in Viper~\cite{juhasz2014viper}, which means they rely on the Rust borrow
checker for lifetime inference but do not need to trust its results.
Doing so, they automate the application of memory reasoning rules and thus avoid
general-purpose memory proof search.
Creusot, however, by virtue of its dedicated
encoding that directly leverages lifetime information, appears to offer better
verification performance than the Prusti frontend for the general-purpose Viper
tool~\cite[§5.3]{creusot}.

\iflong
Our presentation of a low-level language equipped with a system of permissions, followed by an
embedding into a theorem prover, is reminiscent of
RefinedC~\cite{sammler2021refinedc}. RefinedC relies on magic wands to make up
for the lack of borrows; wands, by virtue of being very general, require the use
of heuristics.
RefinedC, however, focuses on the subset of C code that obeys its permission discipline; and it
relies on a memory model in Coq rather than a functional translation. RefinedC is foundational,
and requires user annotations in an intrinsic style,
while \aeneas generates a trusted, pure translation for extrinsic proofs.
Crucially, both RefinedC and \aeneas rely on the fact that they never need to
backtrack after applying rules. For RefinedC, this is made possible by restricting the
user annotations to a carefully crafted, yet extensible, fragment of separation logic, along with
alias types in the style of Mezzo's permissions~\cite{protzenko2014mezzo}.
For Aeneas, we never backtrack because we leverage the borrow mechanism to lazily terminate
borrows.

\aeneas is inspired by the Mezzo language~\cite{pottier2013programming}.
In Mezzo, the type-checker is flow-sensitive and operates by maintaining a permission environment
for each program point, possibly losing information for function calls and disjunctions in the
control-flow. This design is similar to our symbolic semantics. This also further illuminates our
earlier claim, which is that our symbolic interpreter acts as a borrow-checker, and further
encourages us to investigate a standalone Rust borrow-checker based on the \aeneas formalism.

Heapster~\cite{heapster}, just like \aeneas, attempts to
extract pure code from a low-level program. In the case of Heapster, the input is LLVM internal
code, and the output is logical (non-executable) specifications.
The user must guide extraction by adding type annotations and loop invariants
to their programs. A key difference lies in the way
they handle pointers to elements inside recursive data-structures,
a typical use case for borrows in Rust, by requiring
the user to define what they dub \emph{reachability permissions}.
Those permissions act as predicates describing the link between a pointer
and the value it points to inside a structure, and allow them to reason about the
effect of destructive updates. Reasoning-wise, reachability permissions fulfill
a role similar to our backward functions.
With \aeneas, however, we leverage Rust's borrow discipline to automate
the generation of those functions.
\fi

Cogent~\cite{cogent,cogentold} is a domain-specific language equipped with a linear type
system. The Cogent compiler produces: C code; a high-level Isabelle/HOL
specification; and a proof of refinement from the former to the latter.
By virtue of producing an Isabelle/HOL specification, Cogent seamlessly composes
with existing developments in that language, and can thus be integrated into a
larger project, something \aeneas also enables. However, unlike \aeneas, the
Cogent compiler does not need to be trusted since it produces a proof of
translation correctness for each compilation run. We also remark that the linear
type system of Cogent is significantly less expressive than Rust's; notably,
Cogent does not seem to allow an equivalent of mutable borrows.

Stacked Borrows~\cite{jung2019stacked} give a semantics to the notion of borrows
in Rust, but sets out to achieve different goals than Aeneas: namely, to provide
a set of rules that Rust developers can follow and validate their code against
when writing unsafe code. The work comes with an extensive evaluation, which
establishes both that the tool can detect incorrect uses (bugs were found), and
that it can prove that some optimizations written using unsafe code are correct.
This work adopts a very low-level view of memory, and it is unclear whether it
can be used productively at the scale that we envision for \aeneas. The value of
the work, however, lies in its precise, memory-based semantics of borrows; we
are evaluating the feasibility of proving our semantics against it.

\fref{comparison} compares some of the tools above to \aeneas.
If anything, the table reveals that each tool adopts a unique stance on what
kind of programs they aim to verify, and with what kind of toolchain. For \aeneas,
the stance is as follows: we target safe Rust programs, and we believe in
extrinsic reasoning. Doing so, we hit what we believe is a ``sweet spot'', where
the functional encoding is lightweight and accessible, and where proof engineers
can be most productive.

\begin{acks}
  We are very grateful to Aymeric Fromherz who bravely proofread this paper
  repeatedly at undue hours, and gave many useful remarks and feedback.
  We warmly thank Xavier Denis and Jacques-Henri Jourdan for insightful
  discussions and useful advice throughout the design of \aeneas.
  We also thank Chris Hawblitzel and Andrea Lattuada for a thorough walk-through
  of Verus and its approach to handling borrow termination. Finally, we thank Ralf
  Jung for many insightful remarks on an early version of this paper.
\end{acks}

\bibliographystyle{ACM-Reference-Format}
\bibliography{paper.bib}

\end{document}